\documentclass[a4paper,12pt,preprint,aps,floatfix]{revtex4}
\usepackage{longtable}
\usepackage{natbib}
\usepackage{graphicx}     
\usepackage{lscape}
\usepackage{color}
\usepackage{amsmath}
\usepackage{algorithm}
\usepackage{float}

\newcommand{\etal}{{\it et al.} }

\newcommand{\Schrodinger}{Schr\"{o}dinger }\newcommand{\cm}{$\rm{cm}^{-1}$}
\newcommand{\Art}{$\rm{Ar}_{2}$}
\newcommand{\reff}{$r_{\rm{eff}}$}

\def\a0{{$a_{\rm 0}$}}

\begin{document}
\hspace{3cm}
\title{Low temperature scattering with the R-matrix method: argon-argon scattering}
\author{
Tom Rivlin$^{1}$\footnote{t.rivlin@ucl.ac.uk}, Laura K. McKemmish$^{1,2}$, K. Eryn Spinlove$^{1}$, Jonathan Tennyson$^{1}$\footnote{j.tennyson@ucl.ac.uk}}
\affiliation{$^{1}$ Department of Physics and Astronomy, University College London, London, WC1E 6BT, UK}
\affiliation{$^{2}$ Department of Chemistry, University of New South Wales, Australia
}
\label{firstpage}

\date{\today}

\begin{abstract}

  Results for elastic atom-atom scattering are obtained as a first practical application of  RmatReact, a new code for generating high-accuracy scattering observables from potential energy curves. RmatReact has been created in response to new experimental methods which have paved the way for the routine production of ultracold ($\mu$K) atoms and molecules, and hence the experimental study of chemical reactions involving only a small number of partial waves.  Elastic scattering between argon atoms is studied here. There is an unresolved discrepancy between different \Art\ potential energy curves which give different numbers of vibrational bound states and different scattering lengths for the \Art\ dimer.  Depending on the number of bound states, the scattering length is either large and positive or large and negative. Scattering observables, specifically the scattering length, effective range, and partial and total cross-sections, are computed at low collision energies and compared to previous results. In general, good agreement is obtained, although our full scattering treatment yields resonances which are slightly lower in energy and narrower than previous determinations
using the same potential energy curve.

\pacs{Valid PACS appear here}

\end{abstract}

\maketitle

\section{Introduction}
\label{Introduction}

Laser cooling, Stark deceleration, buffer-gas cooling, and evaporative cooling 
are among a wide variety of cooling techniques developed in recent decades, 
which have allowed for precise control over individual molecules, especially 
diatomic molecules
\cite{14StHuYe.Rmat,12StHuYe.Rmat,10ShBaDe.Rmat,14ZhCoWa.Rmat}. As such, a 
variety of experiments at millikelvin and microkelvin temperatures (so-called 
`ultracold' temperatures) have now become routine, including experiments that 
probe collisions between small species in unprecedented levels of detail
\cite{14MoGrJi.Rmat,18DaLeTo.Rmat,01Bohn.Rmat,06KoGoJu.Rmat,15HeSo.Rmat,
09BeSo.Rmat,09BeGiOl.Rmat}.
This has led to the possibility of the fine-tuning of state-to-state reaction 
dynamics for reactions involving only a small number of partial waves 
\cite{12QuJu.Rmat,10OsNiWa.Rmat}. To quote Stuhl \etal \cite{14StHuYe.Rmat}, 
this is \lq\lq perhaps the most elementary study possible of scattering and 
reaction dynamics\rq\rq. These experiments have
led to the discovery of a variety of intriguing quantum phenomena, including 
shape resonances, Feshbach resonances 
\cite{10ChGrJu.Rmat,06KoGoJu.Rmat,08PeGaCo.Rmat,09HuBeGo.Rmat},
universal scaling laws \cite{04DISuEs.Rmat,10LoOtSe.Rmat}, and Efimov trimers 
\cite{10LoOtSe.Rmat,11FeZeBe.Rmat,09WaEs.Rmat,06DIEs.Rmat}.

Ultracold experiments have also revealed that, as with the well-known, 
near-dissociation $\rm{H}_3^+$ spectrum 
\cite{82CaBuKe.Rmat,jt185,00KeEuMc.Rmat,jt706}, ultracold atomic collisions can 
have an overwhelmingly large density of resonances in
scattering energy \cite{13MaQuGo.Rmat}. Resonance states also offer the best 
opportunity for quantum control and steering: they are already being used to 
steer the formation of ultracold diatomic molecules \cite{14MoGrJi.Rmat}. 

In response to these developments, this paper demonstrates a novel algorithm for 
the simulation of collisions between atoms, with the intention of extending the 
methods to collisions involving larger systems. This algorithm, known as 
RmatReact, is based on the computable R-matrix-based method widely applied to 
electron-atom and electron-molecule collisions \cite{jt474,11Burke.Rmat}, which 
here has been adapted to the atom-atom case. The method is designed to study 
reactive and non-reactive, and elastic and inelastic collisions occurring over 
deep wells. 
With the exception of a single proof-of-principle study by Bocchetta and Gerratt 
\cite{85BoGe.Rmat}, this method has not been applied to so-called heavy particle 
scattering before. 

The R-matrix method, being time-independent, is well-suited to studying the narrow, short-lived resonances considered here. In contrast to existing methods using only R-matrix propagation for heavy particle collisions \cite{76LiWa.Rmat,80WaLi.Rmat}, the R-matrix method employed 
in this work makes full use of the partitioning of space into inner, outer, and asymptotic regions. This is in order to leverage the efficiency of variational nuclear motion programs at solving the 
short-range (inner region) problem and the R-matrix method for generating high resolution plots of scattering observables such as the cross-section. The R-matrix method is very similar in spirit to the multichannel quantum defect theory (MQDT) which has been extensively used to study ultracold atom-atom \cite{98BuGrBo.MQDT,04RaMi.MQDT,05GaTuWi.MQDT} and atom-molecule collisions \cite{11CrWaHu.MQDT,12CrHuJu.MQDT}. 
Both methods consider the problem in two regions and the treatment of the outer region can be very similar. However, while MQDT approximates a full solution of the close-coupling equations by using quantum defects which only have a weak dependence on the collision energy, the R-matrix method aims to provide an exact solution to the close-coupling problem based on an inner region with no energy dependence.

The RmatReact algorithm developed in this paper has been discussed in two previous papers. In Tennyson \etal \cite{jt643} we presented a preliminary formalism, though the method has evolved since then. In Rivlin \etal 
\cite{jt727} we provided a demonstration of the method with comparisons to analytic Morse potentials.

In this paper, numerical results from this new algorithm are presented for the 
elastic scattering of argon atoms off other argon atoms at ultracold 
temperatures, ranging from sub-$\rm{\mu K}$ temperatures up to approximately 1 K 
($ = 0.695$ \cm , where \cm\ is used as a unit of energy). Much of the work in 
this paper is dedicated to re-creating existing results, and confirming known 
pieces of physics, in order to demonstrate the efficacy of the new methods 
developed as part of RmatReact.

Several different ground state \Art\ potential energy curves (PECs) 
\cite{05PaMuFo.Ar2, 93Aziz.Ar2, 77AzCh.Ar2, 10SoWaWu.Ar2, 03TaTo.Ar2, 
10PaSz.Ar2, 18MyDhCh.Ar2} are examined in order to simulate this scattering. 
These PECs are listed in Table~\ref{tab:PECs}.
Despite having a shallow PEC formed from van der Waals forces, similar to other 
noble gas dimers, and in contrast to the more deep-well systems that the method 
was designed to study \cite{jt643}, the
argon-argon system has been chosen as a test system for the algorithm. This is 
in order to compare against existing experimental and computational results. The 
large number of high-accuracy PECs available for \Art\ make it a good candidate 
for testing the RmatReact method.  Experiments have also been performed on cold 
ground state argon atoms \cite{14EdBa.Ar2}.

\begin{table}
\centering
\caption{The five PECs studied in this work, with their minima and equilibrium distances. The number of $J=0$ bound states derived in this work is the same as in all cited references (see Section \ref{Results})). Note here the differing numbers of calculated bound states ($N_{\rm{bound}}$) between the methods.}
\begin{tabular}{cccccc} 
\hline
\,\,Label\,\, & \,\,Authors\,\, & \,\,Citation\,\,                             & \multicolumn{1}{c}{\,\,$N_{\rm{bound}}$\,\, } & \,\,$V_{\rm{min}}$ / \cm\,\, & \,\,$r_{\rm{min}}$ / \AA\,\,  \\ 
\hline
PM   & \,\,Patkowski \etal \,\, & \cite{05PaMuFo.Ar2} & 9                                           & -99.269                                                  & 3.7673                                                    \\ 
Aziz  & Aziz & \cite{93Aziz.Ar2}   & 8                                           & -99.554                                                  & 3.7570                                                    \\ 
TT   & Tang \etal & \cite{03TaTo.Ar2}   & 8                                           & -99.751                                                  & 3.7565                                                    \\ 
PS   & Patkowski \etal & \cite{10PaSz.Ar2}   & 9                                           & -99.351                                                  & 3.7624                                                    \\ 
MD   & Myatt \etal & \cite{18MyDhCh.Ar2} & 8                                           & -99.490                                                  & 3.7660                                                    \\
\hline
\end{tabular}
\label{tab:PECs}
\end{table}

Barletta \etal \cite{jt486} studied low-energy Ar-Ar collisions in support of 
experimental studies using Ar for sympathetic cooling \cite{jt458}; they 
assessed four PECs for the \Art\ system. Of these four, three are also assessed 
in this work (PM, Aziz, TT, see Table \ref{tab:PECs}); the fourth PEC of 
Slav{\'{\i}}{\v{c}}ek \etal \cite{03SlKaPa.Ar2} is not considered here, but two 
additional ones are. 
All five PECs studied here superficially appear very similar. However, as Table 
\ref{tab:PECs} shows,
they do have slight differences which have significant impacts on their 
low-energy scattering properties. The PM and PS PECs were generated \textit{ab 
initio}, whilst AZ, TT and MD used experimental results in their fit.

Barletta \etal \cite{jt486} generated scattering lengths and effective ranges 
for the PECs they studied. These values, especially the scattering lengths, 
diverge significantly from each other. A further PEC, and the associated 
scattering length prediction (computed with the method of Meshkov \etal 
\cite{11MeStLe.Rmat}), from Myatt \etal \cite{18MyDhCh.Ar2}, is also recreated 
here.

The issue of the highly varying scattering lengths appears to be closely linked 
to a long-standing debate over the number of vibrational ($J=0$) bound
states belonging to the \Art\ system. Some PECs appear to support only eight 
bound states, while others appear to support a ninth bound state. If this
state exists, it has a binding energy on the order of magnitude of 1~$\rm{\mu 
K}$, which is approximately 0.7~$\rm{\mu}$\cm , or 86 picoelectronvolts, and 
thus would be difficult to detect. Nevertheless, the value of the scattering 
length of a particular system is highly dependent on the position of the highest 
bound state \cite{11Burke.Rmat}. Consequently, whether or not this state exists 
has important implications for the physics of the scattering.

Sahraeian \etal \cite{18SaHa.Ar2} study two \Art\ PECs; those which we have labelled PM and PS. They claim to have detected the ninth bound state in both cases. This result for the PM PEC is in agreement with Barletta \etal \cite{jt486}.

The RmatReact method is described in Section \ref{TheRmatReactMethod}. The 
results presented in Section \ref{Results} include predictions of the scattering 
length and effective range, and partial and total cross-sections for a variety 
of partial waves, including the detection and characterisation of three 
ultracold shape resonances. These results are compared against literature 
results. In Section \ref{Conclusions}, along with the conclusions, some 
allusions to intended future works with this
algorithm are presented.

\section{The RmatReact Method}
\label{TheRmatReactMethod}

R-matrix theory has existed in various forms since its invention by Eugene 
Wigner in the 1940s \cite{46Wigner.Rmat,47WiEi.Rmat}. The underlying
principle behind the R-matrix method is the partitioning of space into an inner 
region, an outer region, and an asymptotic region along the reaction coordinate 
$r$ \cite{11Burke.Rmat}. The radius of the
boundary between the inner and outer regions is often designated $a_0$.

Since the reaction is assumed to be spherically symmetric, it can be modelled as 
taking place over one dimension, here represented by the internuclear distance, 
$r$. The angular dependence of the scattering observables is accounted for by 
splitting the overall three-dimensional scattering wavefunction into 
one-dimensional partial waves and summing over these waves. Each partial wave 
is labelled by a different value of $J$, which for the system studied here is 
the total angular momentum of the system.

In the inner region, the reactants are treated as a bound system. For two atoms 
scattering off each other, this means that the inner region consists of a 
one-dimensional diatomic PEC. The RmatReact method
solves the one-dimensional, time-independent \Schrodinger equation for this 
system over a range of $r$ values from a minimum of $r_{\rm{min}}$ to a maximum 
of $a_0$. Because the \Schrodinger equation is being solved over a finite region 
instead of over all space, an extra surface term must be added to the equation 
to account for the surface term in the integration. This is known as the Bloch 
term \cite{69Robson.Rmat}. Note that this method differs from some R-matrix implementations, where a Buttle correction 
\cite{67Buttle.Rmat,11Burke.Rmat} is used to account for this issue.

In solving the time-independent \Schrodinger equation with the Bloch term, the 
method diagonalises the inner region to produce finite-region rovibronic 
eigenenergies and eigenfunctions of the diatomic system, which are needed to 
construct the R-matrix on the boundary $a_0$ \cite{11Burke.Rmat}. The 
calculations in this region are independent of scattering energy, and so can be 
performed once for a given symmetry and for all scattering energies, hence 
greatly reducing the computational expense of the method.

\begin{figure*}[htb!]
\centering
\includegraphics[scale=0.75]{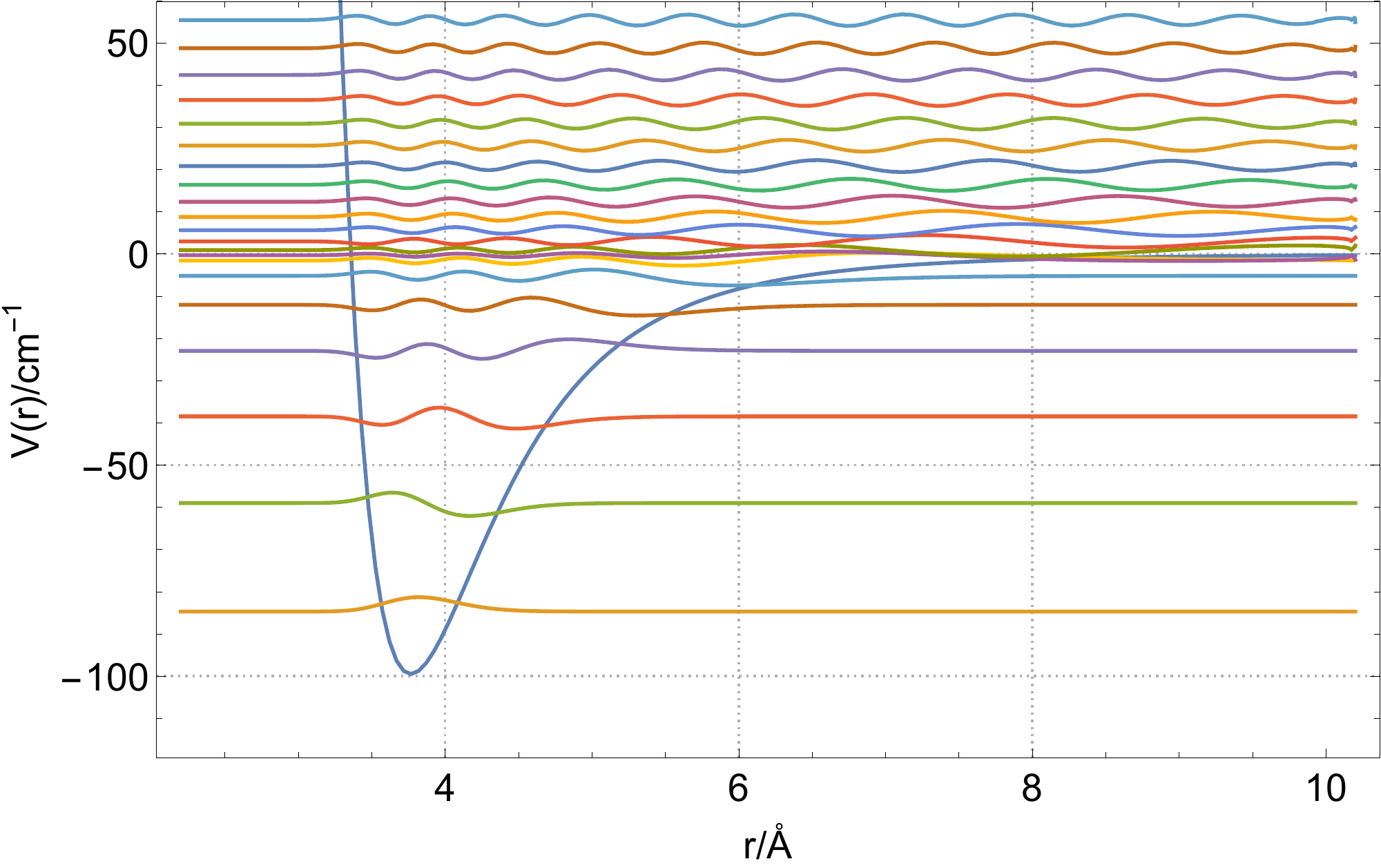}
\caption{Inner region wavefunctions: One of the potential energy curves studied in this work, MD \cite{18MyDhCh.Ar2} with bound and continuum states shown on the curve for an R-matrix inner region ranging from $r_{\rm{min}} = 2.2$~\AA\ to $a_0 = 10.2$~\AA\ and integration over 200 Lobatto grid points. }
\label{fig:Ar2}
\end{figure*}

Figure~\ref{fig:Ar2} shows the PEC of Myatt \etal (MD) \cite{18MyDhCh.Ar2}. The 
eight bound states predicted by Myatt \etal are located below dissociation at 
their appropriate eigenenergies (with two states close to dissociation at 
$-1.58$ \cm\ and $-0.307$ \cm ). The lowest of the continuum states used by the 
RmatReact algorithm in the R-matrix calculation are also shown. At the scale 
shown in Figure~\ref{fig:Ar2}, the MD PEC is not distinguishable from the other 
PECs studied in this work, and the eigenfunctions are very similar.

Slight numerical instability can be seen in the continuum states in the vicinity 
of $a_0$ in Figure~\ref{fig:Ar2}. This is in part due to the smaller number of 
grid points used in the calculation to produce Figure~\ref{fig:Ar2} (as opposed 
to the larger number of grid points used in the results section), and is not 
sufficient to significantly impact the results. In practice, it is only the very last point 
in the grid which is important for the R-matrix calculations.

In the outer region, the reactants are treated as being unbound. However, they 
still interact over a long-range PEC. In this work only PECs which are 
polynomial in $r^{-1}$ at large values of $r$ are considered. The boundary 
between the outer and asymptotic regions is denoted here as $a_p$. In the outer 
region, the RmatReact method uses R-matrix propagation techniques, such as those 
due to Light and Walker \cite{76LiWa.Rmat} or Baluja \etal \cite{82BaBuMo.Rmat} 
to extract the value of the
R-matrix function at $a_p$ from its value at $a_0$. Although this propagation 
method does depend on scattering energy, it is considerably less expensive than 
the inner region calculation.

In the asymptotic region, the potential is assumed to be zero. Here, the 
scattering observables are calculated by determining the K-matrix at a given 
energy \cite{11Burke.Rmat}. Eventually the RmatReact method will utilise an 
asymptotic expansion, such as those developed by Burke and Schey 
\cite{62BuSc.Rmat} or Gailitis \cite{76Ga.Rmat}, for this calculation. However, 
at this point a simpler K-matrix formulation is used (as described below).

Low-energy resonances often have very narrow widths when plotted as a function 
of scattering energy. There are also often many resonances close together. As 
such, it is important to determine the scattering
observables on a fine grid of energies. The R-matrix method's inner-outer region 
separation is ideal for this task.

The details of the R-matrix method for the single-channel (elastic scattering) 
case have been discussed extensively in our previous RmatReact papers 
\cite{jt643,jt727}, although there are differences from the version of the 
method in \cite{jt643}. The following is an abridged explanation derived from 
Burke's \textit{R-Matrix Theory of Atomic Collisions} \cite{11Burke.Rmat}.

\subsection{R-matrix theory}
\label{Rmatrixtheory}

The R-matrix is a quantity with two equivalent definitions. In the single-channel case, the first definition is:
\begin{equation}
\label{eq:RmatrixInner}
R^J(k,a_0) = \frac{1}{a_0}\sum_{n=1}^{N}\frac{\left(w_n^J(a_0)\right)^2}{\left(k_n^J\right)^2 - k^2},
\end{equation}
where $k$ is the scattering wavenumber associated with the scattering energy $E$ via the equation
\begin{equation}
\label{eq:kdef}
k = \sqrt{\frac{2\mu E}{\hbar^2}},
\end{equation}
$R^J(k,a_0)$ is the R-matrix for a certain partial wave $J$, and $\mu$ is the reduced mass of the system.

$k_n^J$ are the wavenumbers associated with the rovibronic eigenenergies $E_n^J$ 
of the diatomic system in the inner region (labelled by quantum numbers $J$ and 
$n$, and following a similar relationship to Equation~(\ref{eq:kdef})), and 
$w_n^J(a_0)$ are known as boundary amplitudes. $E_n^J$ and $w_n^J(a_0)$ are the 
eigenvalues and eigenfunctions (evaluated at $a_0$) respectively of the 
time-independent \Schrodinger equation with the Bloch term: 
\begin{equation}
\left(\hat{H}^J + \mathcal{L}(a_0)\right) w_n^J(r) = E_n^J w_n^J(r),
\label{eq:TISE}
\end{equation}
where $\hat{H}^J$ is the Hamiltonian for the system for a given $J$, which 
includes kinetic and potential operator components, and the Bloch term, 
$\mathcal{L}(a_0)$, is given by
\begin{equation}
\mathcal{L}(a_0) = \delta \left(r - a_0\right) \frac{d}{dr}.
\label{eq:Bloch}
\end{equation}

Burke \cite{11Burke.Rmat} also provides a detailed explanation of how to derive the first definition of the R-matrix given in Equation~(\ref{eq:RmatrixInner}) from the \Schrodinger equation with the Bloch term.

The eigenenergies and eigenfunctions of Equation~(\ref{eq:TISE}) are not restricted to bound states (in fact bound states tend to contribute little to the R-matrix sum). The numerical diagonalisation method used in the inner region creates a discretised continuum of $N - N_{\rm{bound}}$ above-dissociation states, see Figure~\ref{fig:Ar2}, which all contribute to the R-matrix sum.

Note that this definition is entirely dependent on parameters which appear in the inner region and arise from the bound diatomic problem. In contrast, the second definition can be written as
\begin{equation}
\label{eq:RmatrixOuter}
F^J(k,a) = a R^J(k,a)\frac{dF^J(k,r)}{dr}\biggr\rvert_{r=a},
\end{equation}
where $F^J(k,a)$ is the wavefunction associated with a particular partial wave 
$J$, which is evaluated at a particular point $a$. In the multichannel case, 
$F_i^J(k,a)$ is associated with a particular atomic channel $i$. Hence it is 
known as a \textit{channel function}. $R^J(k,a)$ is the R-matrix, as in 
Equation~(\ref{eq:RmatrixInner}). This definition is based on quantities that 
exist in the outer region. From this definition one can see that the R-matrix 
can be thought of as a form of `log-derivative' of the channel function.
This definition of the channel function is consistent with the definition used 
in MQDT \cite{01Gao.Rmat,05GaTiWi.Rmat}, 
which also has similar definitions for the K-matrix and S-matrix (introduced 
below). 

As a result of the equivalence of the two definitions of the R-matrix given by  Equation~(\ref{eq:RmatrixInner}) and Equation~(\ref{eq:RmatrixOuter}), information 
about the inner region bound problem can be used to obtain information about the 
scattering channel functions in the outer region. From these channel functions, 
scattering observables can be constructed via the K-matrix, $K^J(k)$, which is 
dependent on the asymptotic boundary condition involving the channel functions 
at arbitrarily large distances:
\begin{equation}
\label{eq:Kmatrix1}
F_i^J(k,r) \underset{r \rightarrow \infty}{\sim} s_i^J(kr) + K^J(k) c_i^J(kr).
\end{equation}
Here $s_i^J(kr)$ and $c_i^J(kr)$ are `sine-like' and `cosine-like' functions which, in general, can have a variety of forms depending on the specific asymptotic region implementation of the R-matrix method
being used (as described below).

\subsection{Implementation}
\label{Implementation}

The RmatReact method is designed to act as a `harness' between other codes that solve the inner and outer region problems. Ultimately it is intended for the harness to function with a variety of inner and outer region codes, with a \lq\lq plug and play\rq\rq mentality in mind.

In this work, the harness is only used with one inner region code: a modified version of the diatomic nuclear motion code \textsc{Duo} \cite{jt609}. The version of \textsc{Duo} used here has been modified to use a discrete variable representation (DVR) basis \cite{82LiPaLi.Rmat,85LiHaLi.Rmat} based on Lobatto shape functions, and to solve the inner region problem with the additional Bloch term. The Lobatto functions are derived from work by Manolopoulos, Wyatt, and others \cite{93Manolopoulos.Rmat,88MaWy.Rmat,94Meyer.Rmat,89MaWy.Rmat}, which explain how to derive expressions for the kinetic and potential components of the Hamiltonian. 

This is in contrast to the `sinc DVR' method \cite{92CoMi.Rmat} currently 
implemented in \textsc{Duo}, which enforces a zero boundary condition on its 
eigenfunctions at the ends of the grid -- clearly an unacceptable property for a 
method which relies on the amplitudes of eigenfunctions at the boundary. The 
Lobatto DVR method has boundary conditions that set the derivatives of the 
eigenfunctions at the boundary to zero, but the amplitudes themselves are 
allowed to take on non-zero values at $a_0$.
This is also in contrast to the method of Bocchetta and Gerratt 
\cite{85BoGe.Rmat}, which used non-orthogonality and a grid that extended 
slightly beyond $a_0$ to produce arbitrary boundary conditions at $a_0$. This 
approach was tested in earlier versions of this work (see \cite{jt643}), but 
has since been supplanted by the Lobatto DVR methods.

The algorithm for generating Lobatto
shape function nodes and weights is derived from Manolopoulos 
\cite{93Manolopoulos.Rmat}, with some modifications. The \textsc{Duo} code with 
Lobatto functionality used in this work is provided on the
\textsc{Duo} GitHub page.

In this work, the outer region is handled in the harness code itself, with an 
iteration method in space based on the R-matrix propagation methods of Light and 
Walker \cite{11Burke.Rmat,76LiWa.Rmat,76ZvLi.Rmat,79ScWa.Rmat}. In future work, 
however, this will be replaced with the fast R-matrix propagation code PFARM 
\cite{02SuNoBu.Rmat}, based on the sector diagonalisation method of Baluja \etal 
\cite{82BaBuMo.Rmat}, which also includes the asymptotic expansion of Gailitis 
\cite{76Ga.Rmat}. Preliminary testing with PFARM has demonstrated that it is 
able to re-create the resonances described in the results section of this paper.
The specific implementation of the Light-Walker propagator used in this work can 
be seen in Equation~(3) and Equation~(4) of \cite{jt727}.

The asymptotic region is addressed in this work using the following expression for the K-matrix:
\begin{equation}
\label{eq:Kmatrix2}
K^J(k) = \frac{R^J(k,a_p) ka_p s_J'(ka_p) - s_J(ka_p)}{c_J(ka_p) - R^J(E,a_p)ka_p c_J'(ka_p)},
\end{equation}
where $E$ is the scattering energy as before, and where $s_J(ka_p)$ and $c_J(ka_p)$ are given by:
\begin{equation}
\begin{aligned}
\label{eq:sJcJ}
s_J(kr) &= kr j_J(kr)\\
c_J(kr) &= -kr n_J(kr).
\end{aligned}
\end{equation}
Here $j_J(kr)$ and $n_J(kr)$ are the spherical Bessel and Neumann
functions respectively \cite{14GrRy.Rmat}, and the derivatives of
$s_J(kr)$ and $c_J(kr)$ with respect to $r$, at the point $a_p$, are
defined as $s_J'(ka_p)$ and $c_J'(ka_p)$. The distance, $a_p$, should
be chosen such that the potential is sufficiently small by that point
that the value of the K-matrix is not affected by the specific choice
of $a_p$.

\subsection{Scattering observables}
\label{Scatteringobservables}

The four quantities generated by the RmatReact method which are
presented in this work are the eigenphase, cross-section, scattering
length, and effective range of the argon-argon interaction. All of
these can be constructed from the K-matrix, $K(k)$. In the
single-channel case, they have simplified forms
\cite{11Burke.Rmat}. 
The eigenphase, $\delta(k)$, sometimes known as the phase shift, is given by:
\begin{equation}
\label{eq:EP}
\delta(k) = \arctan{K(k)}.
\end{equation}
As a result of this definition, the eigenphase (in radians) is the same
modulo $\pi$. The eigenphases presented in this work are given in the range $[-\frac{\pi}{2},\frac{\pi}{2}]$. This leads to seeming
discontinuities, e.g. in Figure~\ref{fig:EPXS} when the eigenphase passes through $|\frac{\pi}{2}|$. These discontinuities
 are  characteristic of resonances which are also present
in the eigenphases. Although the eigenphase is technically not a
scattering observable itself, it can be used to construct the other
three observables used in this work, and is consequently the best variable for the detection of resonances.

The total cross-section for an interaction can be given as the sum over partial waves from a minimum $J$ value $J_{\rm{min}}$ to a maximum $J_{\rm{max}}$, $\sigma_{\rm{tot}}(k)$. It is given by:
\begin{equation}
\label{eq:XS}
\sigma_{\rm{tot}}(k) =  \sum_{J=J_{\rm{min}}}^{J_{\rm{max}}} \frac{4\pi}{k^2} (2J+1) \sin{(\delta^J(k))}^2,
\end{equation}
and the cross-section for a given partial wave, $\sigma^J(k)$ is merely the summand of Equation~(\ref{eq:XS}).

The scattering length, $A$, and the effective range, \reff , are numbers that characterise the properties of the PEC and the scattering process at low energy. They can be defined in terms of a linear expansion at low energy. If one plots $k \cot{\delta(k)}$ for $J=0$ as a function of $k^2$, then for sufficiently low energy the plot should be linear. In this case, $A$ and \reff\ are defined in the following way:
\begin{equation}
\label{eq:lowenerg}
k \cot{\delta(k)} = -\frac{1}{A} + \frac{1}{2} r_{\rm{eff}} k^2,
\end{equation}
ignoring higher order terms in $k^2$.

Another observable it is possible to detect using the eigenphase is a resonance. 
A resonance will appear as a feature in a plot of the eigenphase or 
cross-section as a function of $E$. Furthermore, the energy of the resonance, 
and its width -- the inverse of its lifetime -- can be determined by fitting a 
function to the eigenphase following the form of Breit and Wigner 
\cite{36BrWi.Rmat,11Burke.Rmat,jt31}:
\begin{equation}
\delta^J(E)= A_0 + A_1 E  + \arctan{\frac{\Gamma_{\rm{res}}}{E-E_{\rm{res}}}} ,
\label{eq:BreitWigner}
\end{equation}
where $\delta^J(E)$ is the eigenphase for partial wave $J$ at scattering energy 
$E$, $\Gamma_{\rm{res}}$ is the width of the given resonance, and $E_{\rm{res}}$ 
is the energy of the resonance. Note this definition of $\Gamma_{\rm{res}}$ 
follows that in standard use in scattering (eg \cite{jt31}), and differs by factor of two from the 
definition a full width at half maximum (FWHM).

The non-resonant shape of the eigenphase (the `background' eigenphase) is 
accounted for by the two terms $A_0$ and $A_1$, where it is assumed that the 
width is narrow enough that the background eigenphase can be approximated by a 
linear function of $E$ over its length. 
By fitting a generated eigenphase to a function of this form, values for 
$\Gamma_{\rm{res}}$ and $E_{\rm{res}}$ can be obtained. Note that it is 
sometimes necessary to replace 
the final term in Equation~(\ref{eq:BreitWigner}) with its negative, if it is 
required by the resonance shape.

\section{Results}
\label{Results}

\subsection{Bound states}
\label{Boundstates}

When performing the inner region calculations with \textsc{Duo} in this work, the number of bound states was found to be in agreement with literature values \cite{jt486,18SaHa.Ar2} for all five PECs studied (see Table \ref{tab:PECs}).

\begin{figure*}[htb!]
\centering
\includegraphics[scale=0.6]{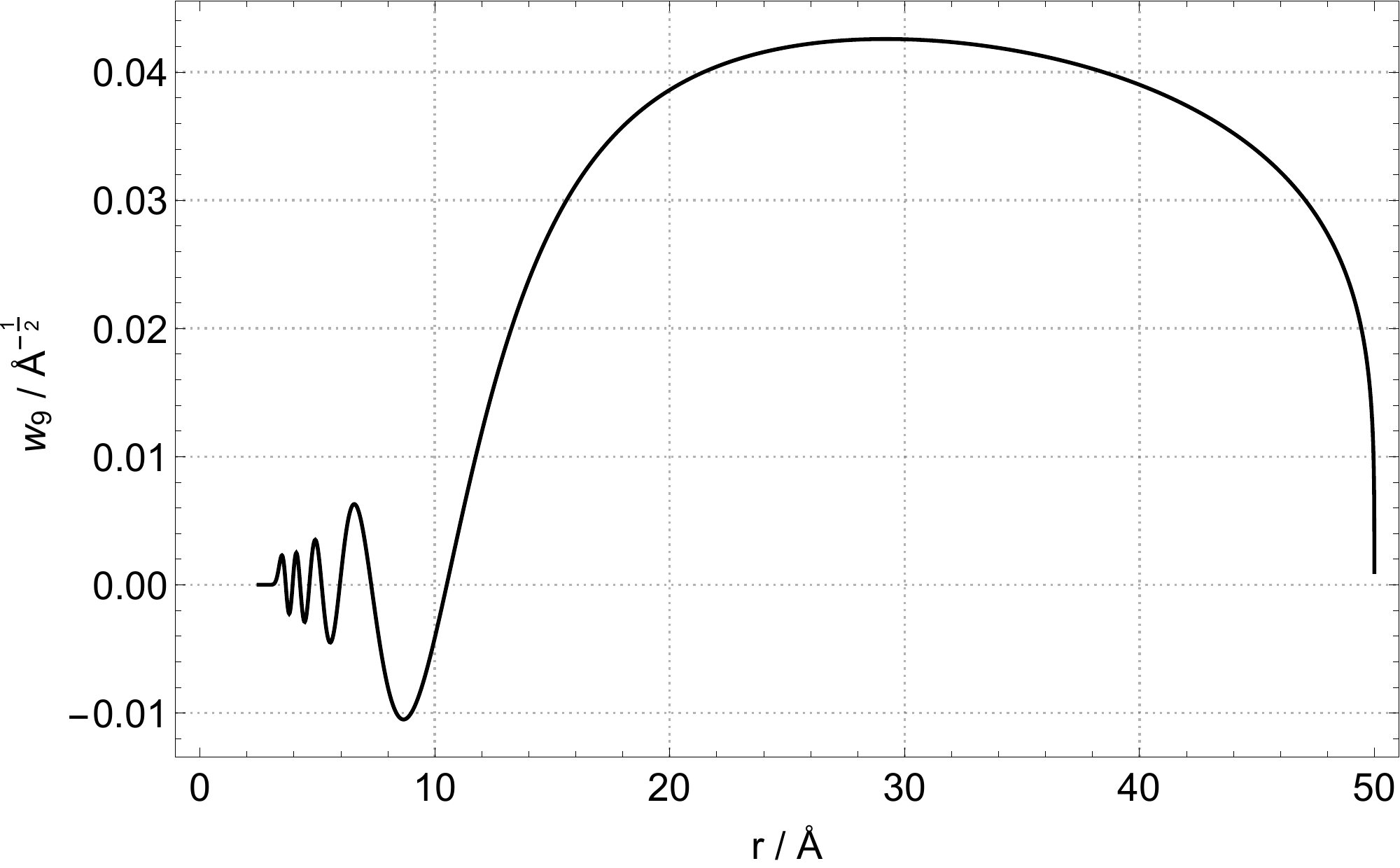}
\caption{The ninth bound state of the PM PEC, plotted as a function of $r$, for when $a_0 = 50$ \AA .}
\label{fig:Psi9PM}
\end{figure*}

However, there were considerable complications when attempting to detect the ninth bound state in this work for the PECs where it was predicted to exist -- the PS and PM potentials. As this state is so weakly bound, it was necessary to extend the inner region calculations out to large distances in order to detect it. This ninth bound state has many similarities to a halo state \cite{18OwSp.Rmat}, as seen in Figure~\ref{fig:Psi9PM}, which shows the ninth bound state as a function of $r$ for the PM PEC for when $a_0 = 50$ \AA .

The results in Table \ref{tab:BarlettaTable} of this work were obtained only by extending $a_0$ out to distances of over 35 \AA\ for the PM PEC and over 40 \AA\ for the PS PEC. As such, a very large number of points needed to be used in 
order to maintain precision. 
The difficulty in detecting the ninth bound state is underlined by the fact that when the diatomic nuclear motion code \textsc{LEVEL} \cite{16LeRoy.Rmat} was used, the ninth bound state was never detected for any of PECs considered here, no matter how far out or how many points the inner region was integrated over. Sahraeian \etal \cite{18SaHa.Ar2} also cited difficulties in detecting this state, which they quote a value of $-0.86233\ \mu$\cm\ for.

\begin{figure*}[htb!]
\centering
\includegraphics[scale=0.6]{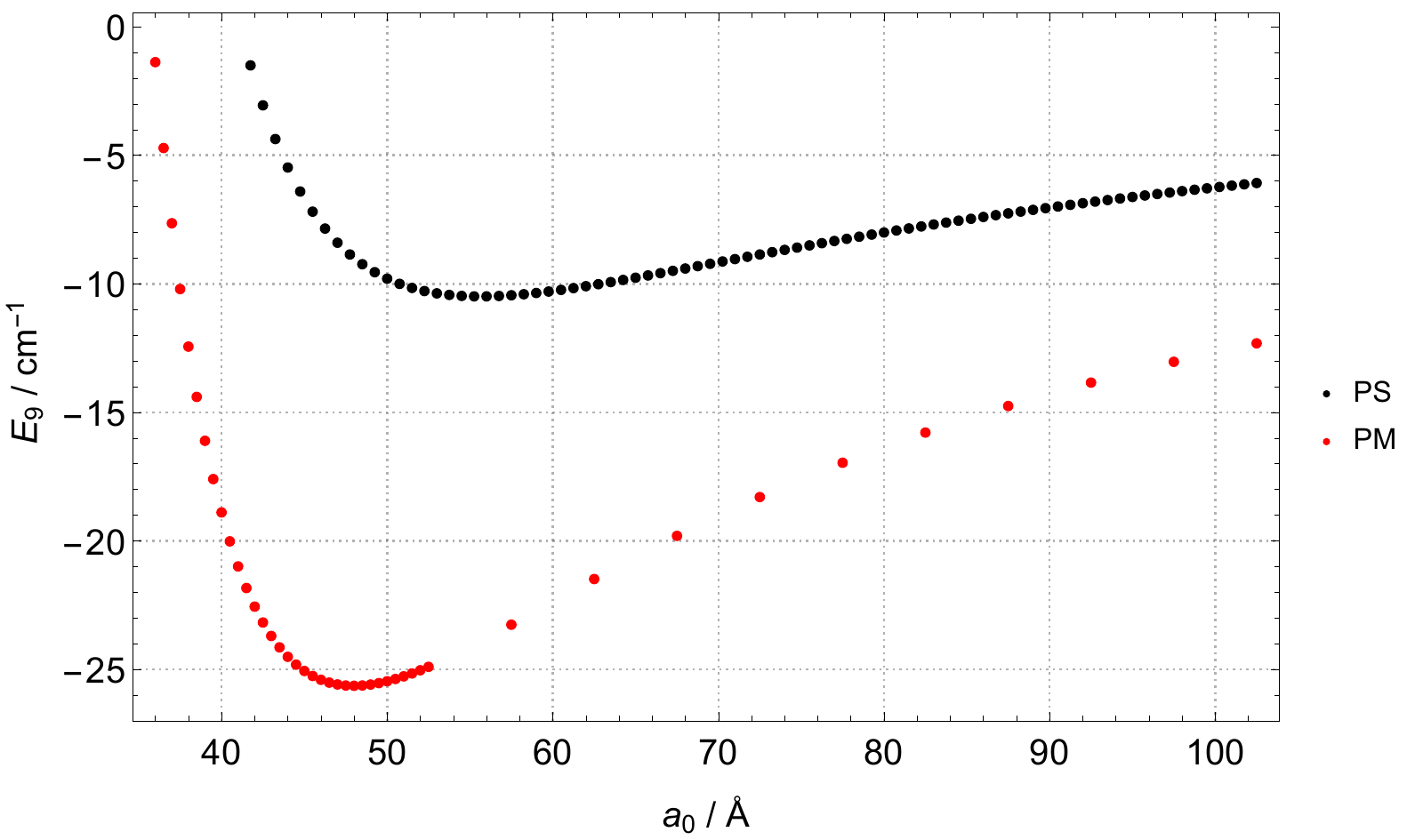}
\caption{The ninth bound state of the PM and PS PECs, plotted as a function of the $a_0$ used in the calculation to generate them, whilst keeping $r_{\rm{min}}$ and the average grid spacing used constant.}
\label{fig:E9PM}
\end{figure*}

Consequently the actual binding energy of the ninth bound state, for PECs in which it was 
detected, varied as a function of the $a_0$ used in the integration here, up to 100 \AA . This is seen in Figure~\ref{fig:E9PM}, which shows the value of the ninth bound state, $E_9$, of the PM and PS PECs as a function of $a_0$, for all values of $a_0$ under 105 \AA\ for which the state was actually bound. If the calculations are converging on fixed values of $E_9$, they are significantly different from the value obtained by Sahraeian \etal

Once a threshold value of $a_0$ was reached, every PEC that the literature 
claimed had nine bound states were consistently found to do so, even if its 
value changed with $a_0$. No ninth bound state was detected in this work for any 
PEC for which it was claimed that there are only eight bound states, even when 
using large values of $a_0$ over 100 \AA . More 
assessment of the numerical issues faced by the R-matrix method can be found in 
Rivlin \etal \cite{jt727}.

We note that diffuse bound states whose wavefunctions extend significantly 
beyond $a_0$ can be found rather efficiently within an R-matrix formalism by 
performing scattering calculations with negative energies \cite{jt106,jt560}. We 
plan to implement such a procedure within the RmatReact framework.

\subsection{Resonances}
\label{Resonances}

The supplementary data provided by Myatt \etal \cite{18MyDhCh.Ar2} (MD) includes 
the rovibrational
eigenenergies of the \Art\ system obtained using \textsc{LEVEL} 
\cite{16LeRoy.Rmat}. The supplementary data also quotes values for states which 
lie above the dissociation threshold but below a centrifugal barrier for $J>0$, 
known as quasibound states. The quasibound states from Myatt \etal
\cite{18MyDhCh.Ar2} which have $J$ quantum numbers $J\leq10$ are quoted in Table 
\ref{tab:MyattDharmTable}.

\begin{table}[]
  \caption{Positions ($E_{\rm{res}}$) and widths ($\Gamma_{\rm{res}}$) with standard errors of three quasibound states extracted from the supplementary data of Myatt \etal  (MD) \cite{18MyDhCh.Ar2}, compared to the three shape resonances produced in this work by fitting eigenphases to Equation~(\ref{eq:BreitWigner}) (with background resonance paramters $A_0$ and $A_1$). The widths extracted from Myatt \etal \cite{18MyDhCh.Ar2} have been multiplied by two to match the convention employed in this paper.}
\begin{tabular}{ccclllcc}
\hline
\,\,$v$\,\, & \,\,$J$\,\, &\,\, \begin{tabular}[c]{@{}c@{}}$E_{\rm{res}}$ (MD)\\  /\cm\end{tabular} \,\,&\,\, \begin{tabular}[c]{@{}c@{}}$\Gamma_{\rm{res}}$ (MD)\\  /\cm\end{tabular} \,\,&\,\, \begin{tabular}[c]{@{}c@{}}$E_{\rm{res}}$ (this work)\\  /\cm\end{tabular} \,\,&\,\, \begin{tabular}[c]{@{}c@{}}$\Gamma_{\rm{res}}$ (this work)\\  /\cm\end{tabular} \,\,&\,\, \begin{tabular}[c]{@{}c@{}}$A_0$\\  \,\end{tabular}    \,\,&\,\, \begin{tabular}[c]{@{}c@{}}$A_1$\\  /(\cm)$^{-1}$ \end{tabular}\,\, \\ \hline
6   & 9   & 0.129                                                    & $0.660 \times 10^{-6}$                                     & 0.1287 & $0.663 \times 10^{-6}$                                         & 0        & 0                                                              \\ 
6   & 10  & 0.448                                                    & 0.00330                                                    & 0.4486(2)                                                       & 0.00247(46) & \,\,-0.107\,\, & 0.757\\ 
7   & 5   & 0.071                                                    & 0.00605 & 0.06993(5)                                                         & 0.004841(5)                                                           & \,\,0.00213\,\, & -0.997 \\ \hline
\end{tabular}
\label{tab:MyattDharmTable}
\end{table}

In this work, the quasibound states quoted for the MD potential in
Myatt \etal \cite{18MyDhCh.Ar2} were characterised by analysing
resonances in the scattering calculation. The diatomic nuclear motion
code used in this work, \textsc{Duo}, does not have the capacity to
detect quasibound eigenvalues directly (although it is possible to
detect them using a stabilisation method with continuum states).
However, these quasibound states should correspond
to shape resonances, which can be detected in plots of the
eigenphase and cross-section.

In order to detect the shape resonances, the RmatReact method was used
to generate the eigenphase, and from it the partial cross-sections for all the partial waves with $J\leq 10$. The inner region was
calculated using 500 Lobatto grid points between $r_{\rm{min}} = 2.5$
\AA\ and $a_0 = 22.5$ \AA . The outer region propagation was performed
from $a_0 = 22.5$ \AA\ to $a_p = 45$ \AA , with over 1,000 propagation
iterations.

\begin{figure*}[htb!]
\centering
\includegraphics[scale=0.4]{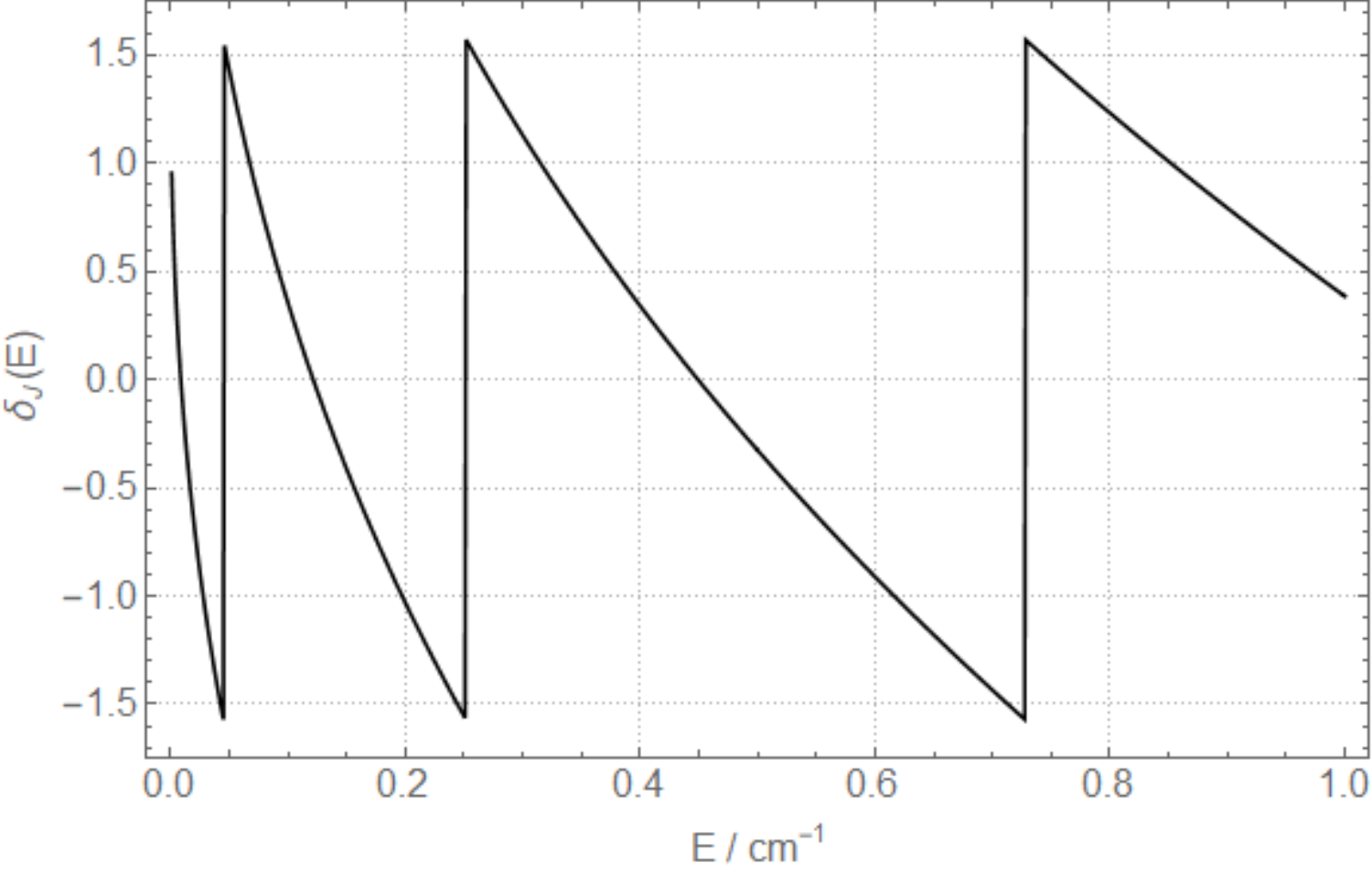}
\includegraphics[scale=0.4]{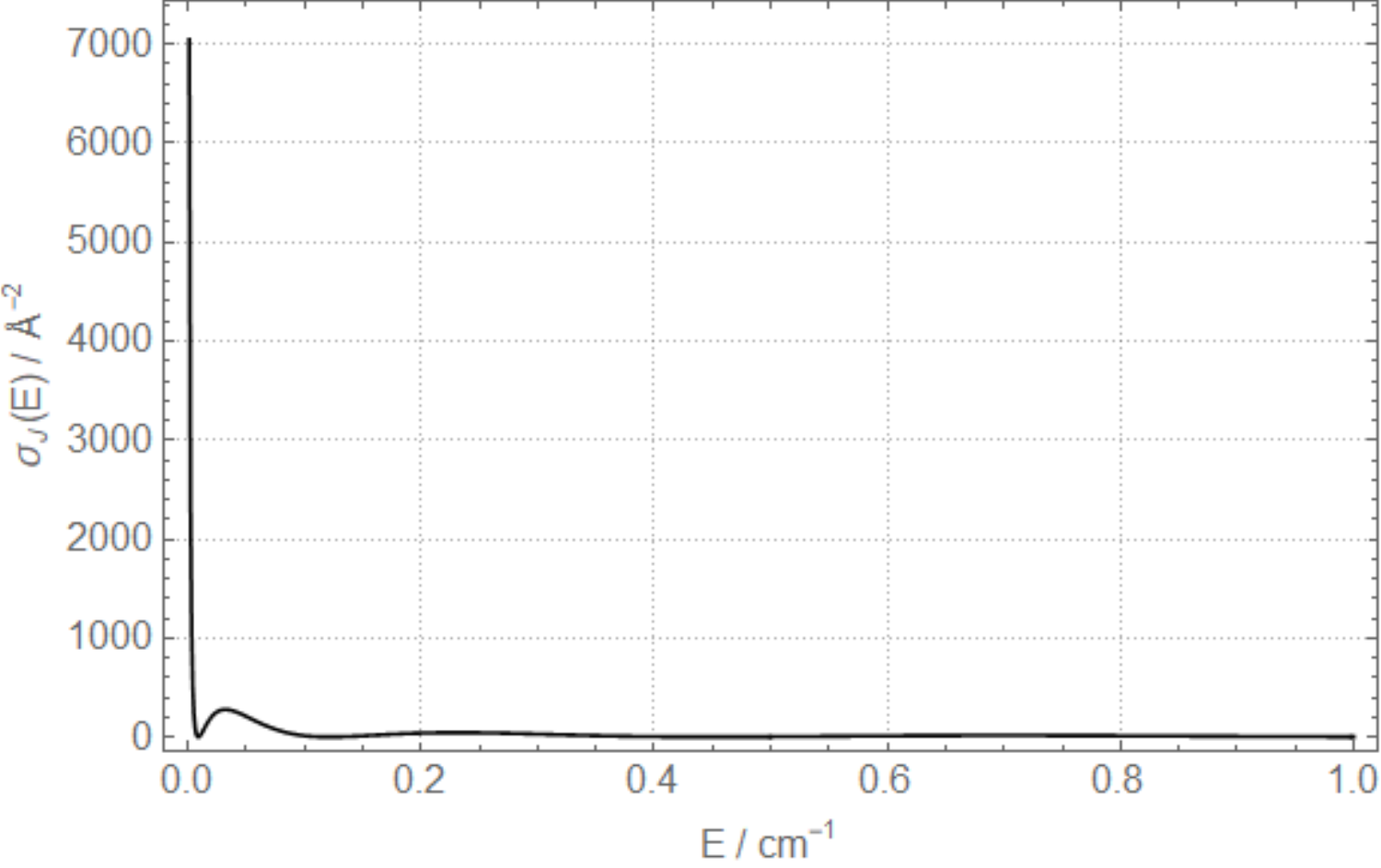}
\includegraphics[scale=0.4]{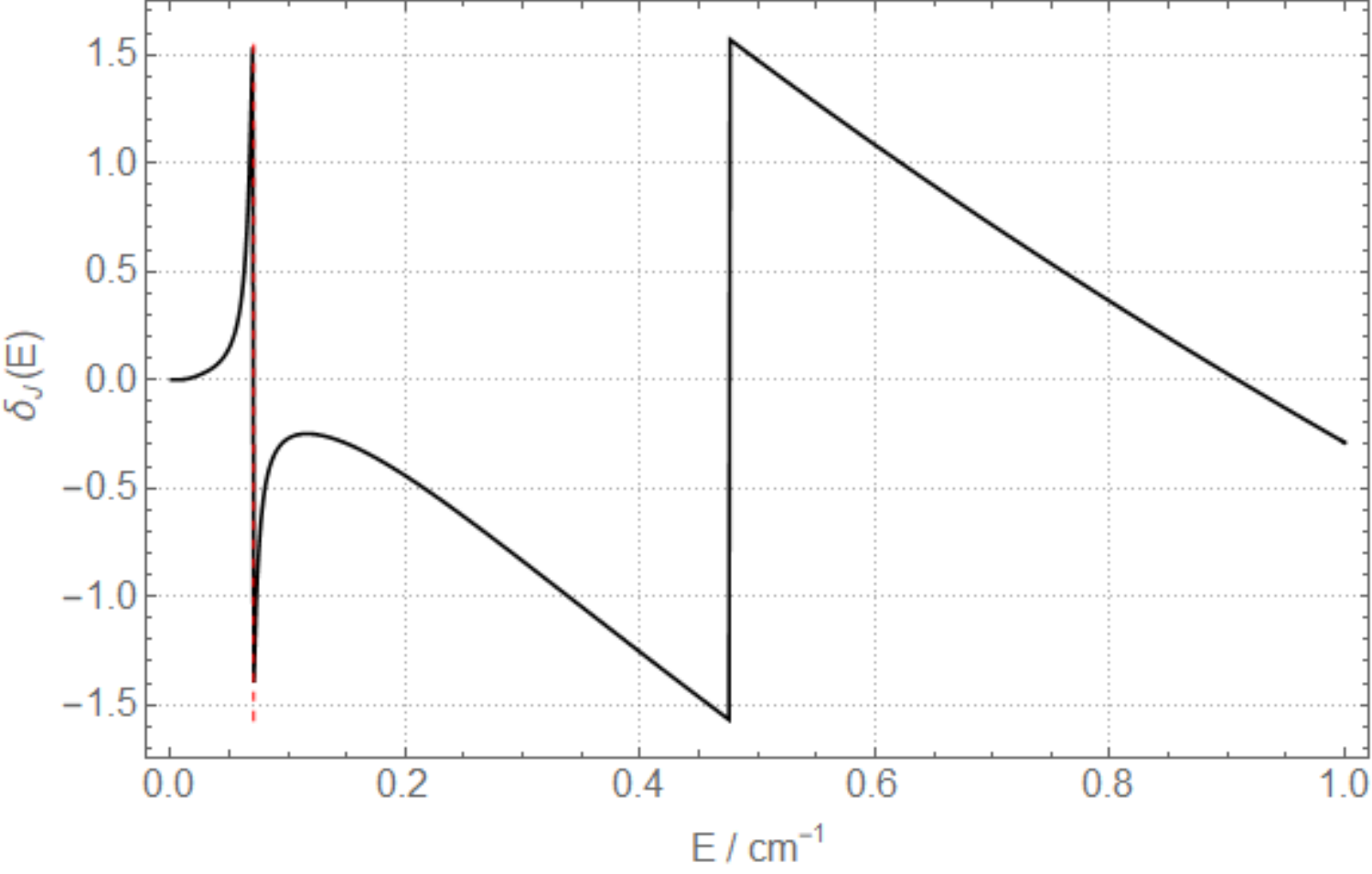}
\includegraphics[scale=0.4]{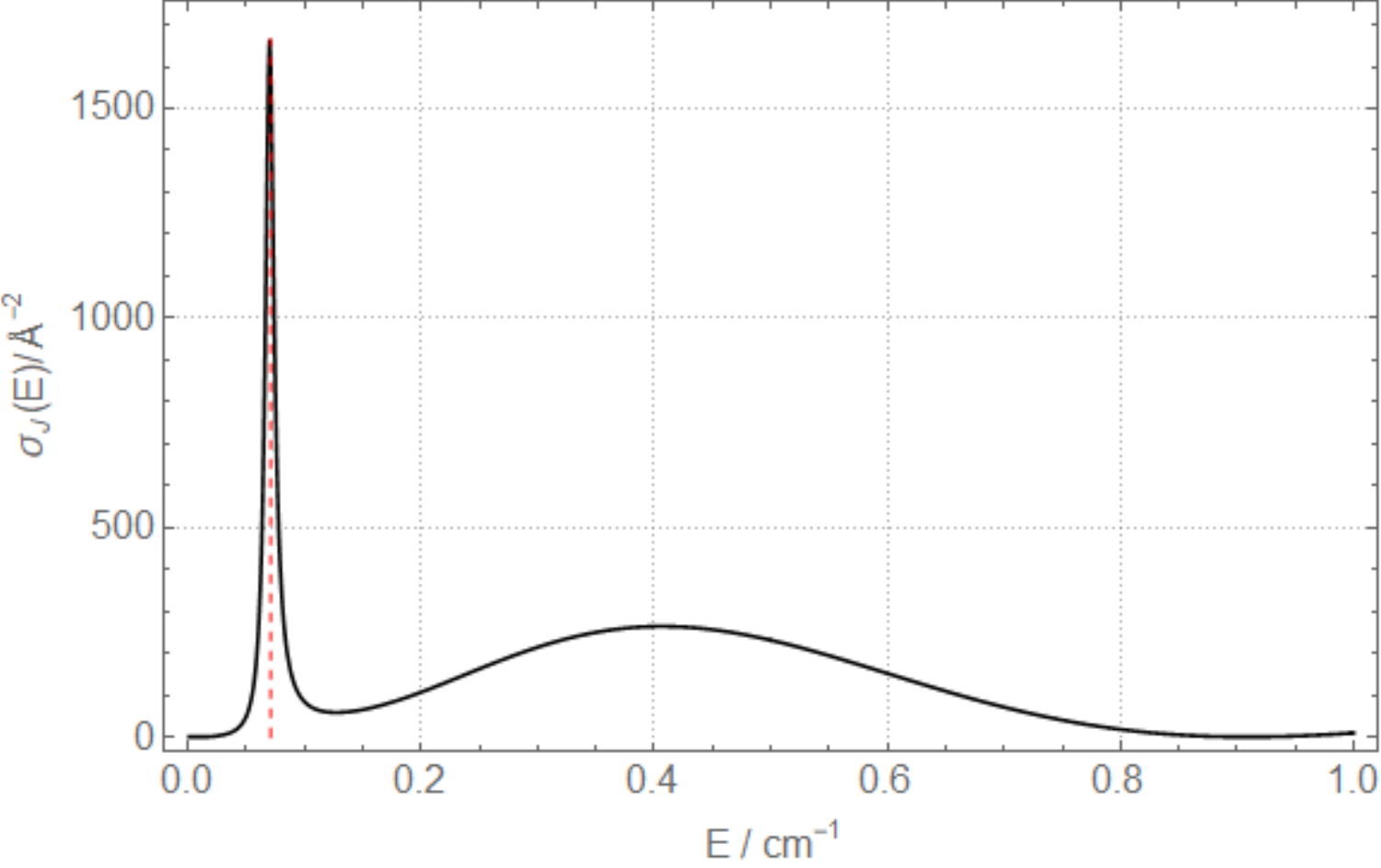}
\includegraphics[scale=0.4]{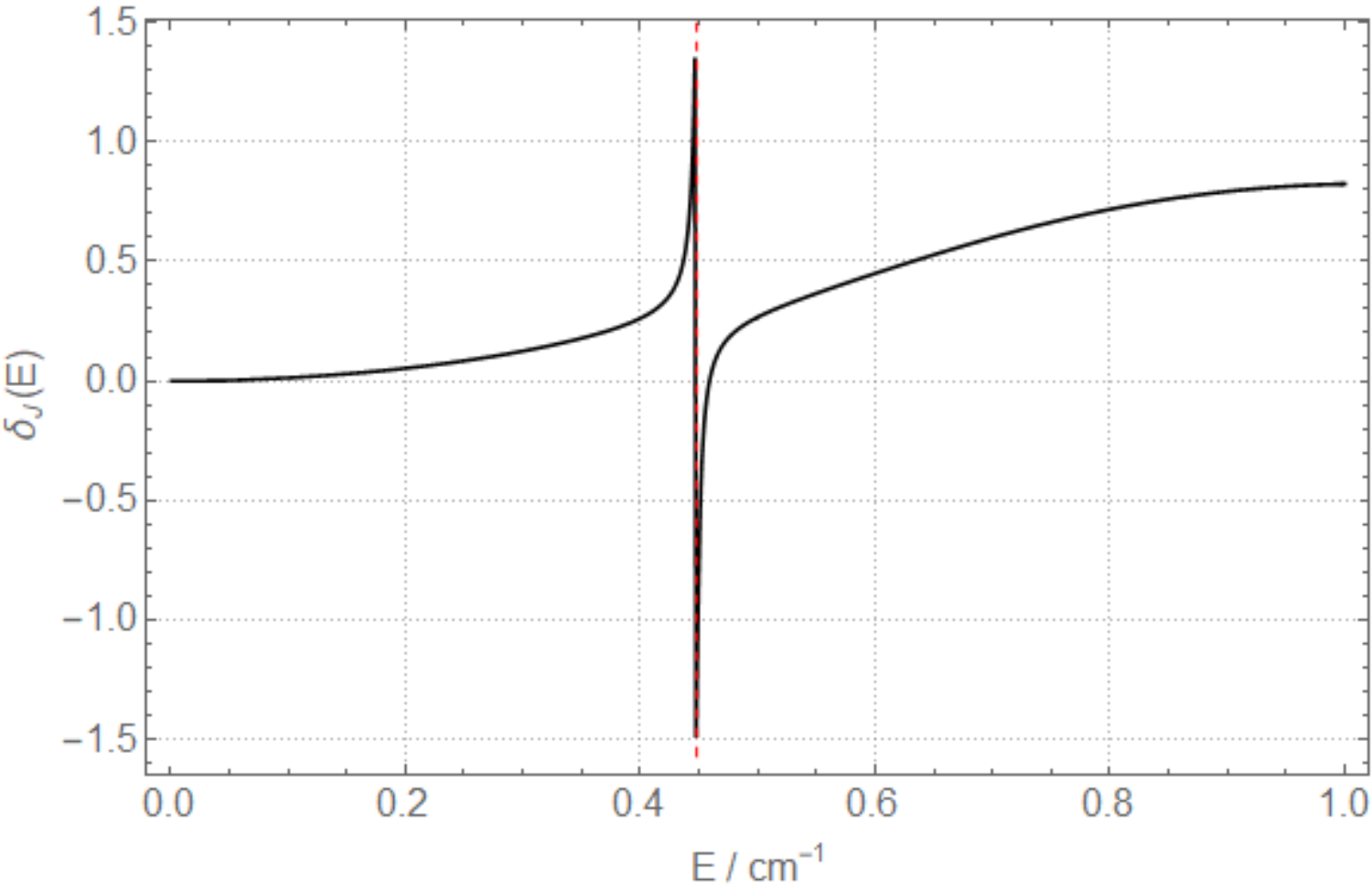}
\includegraphics[scale=0.4]{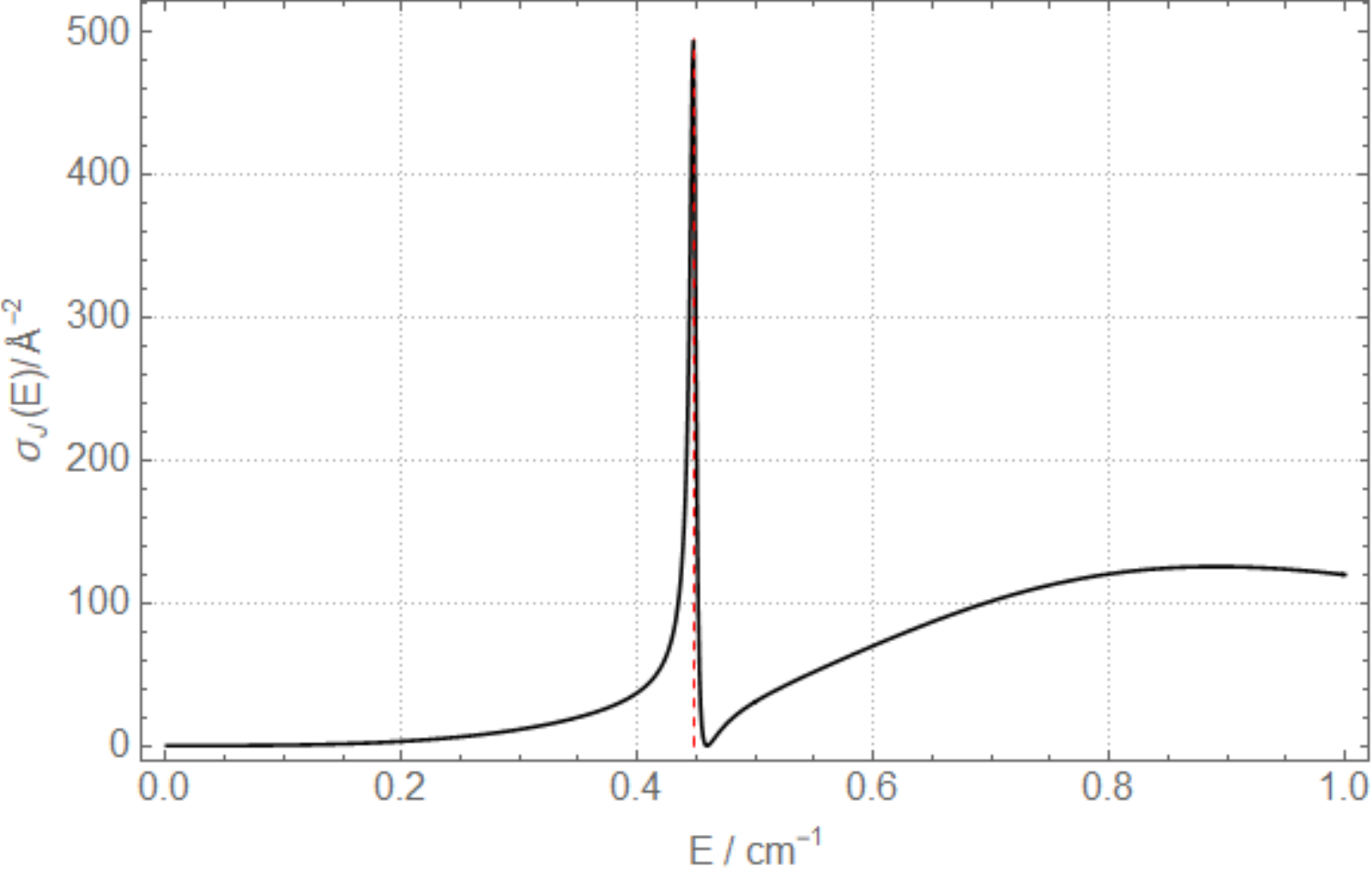}
\caption{Eigenphase (top left, bottom left) and cross-section (top right, bottom right) plots for the $J=0$, $J=5$ and $J=10$ partial waves, generated using the MD potential \cite{18MyDhCh.Ar2}. The dashed red lines mark the position of the resonances. }
\label{fig:EPXS}
\end{figure*}

\begin{figure*}[htb!]
\centering
\includegraphics[scale=0.4]{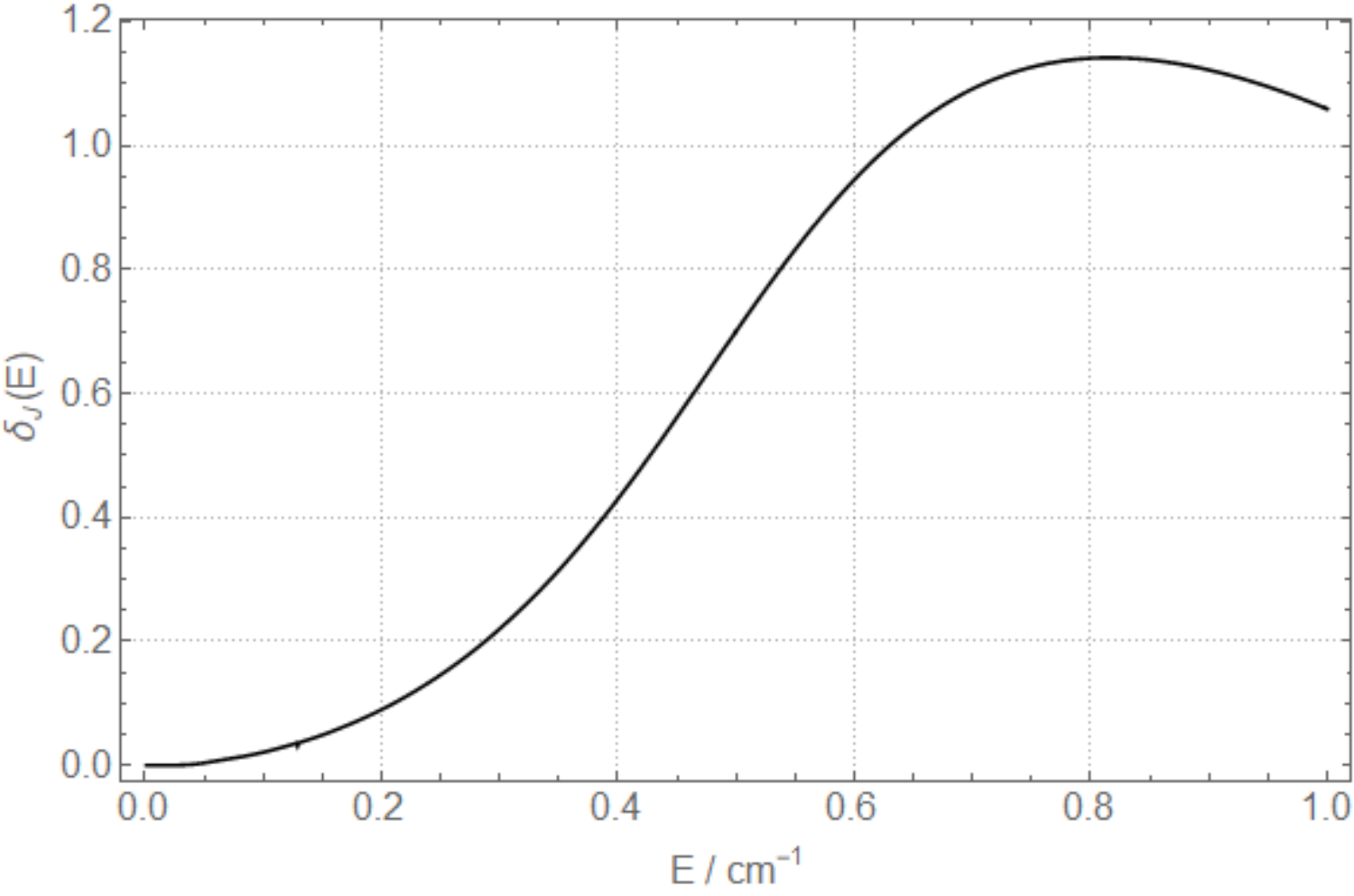}
\includegraphics[scale=0.4]{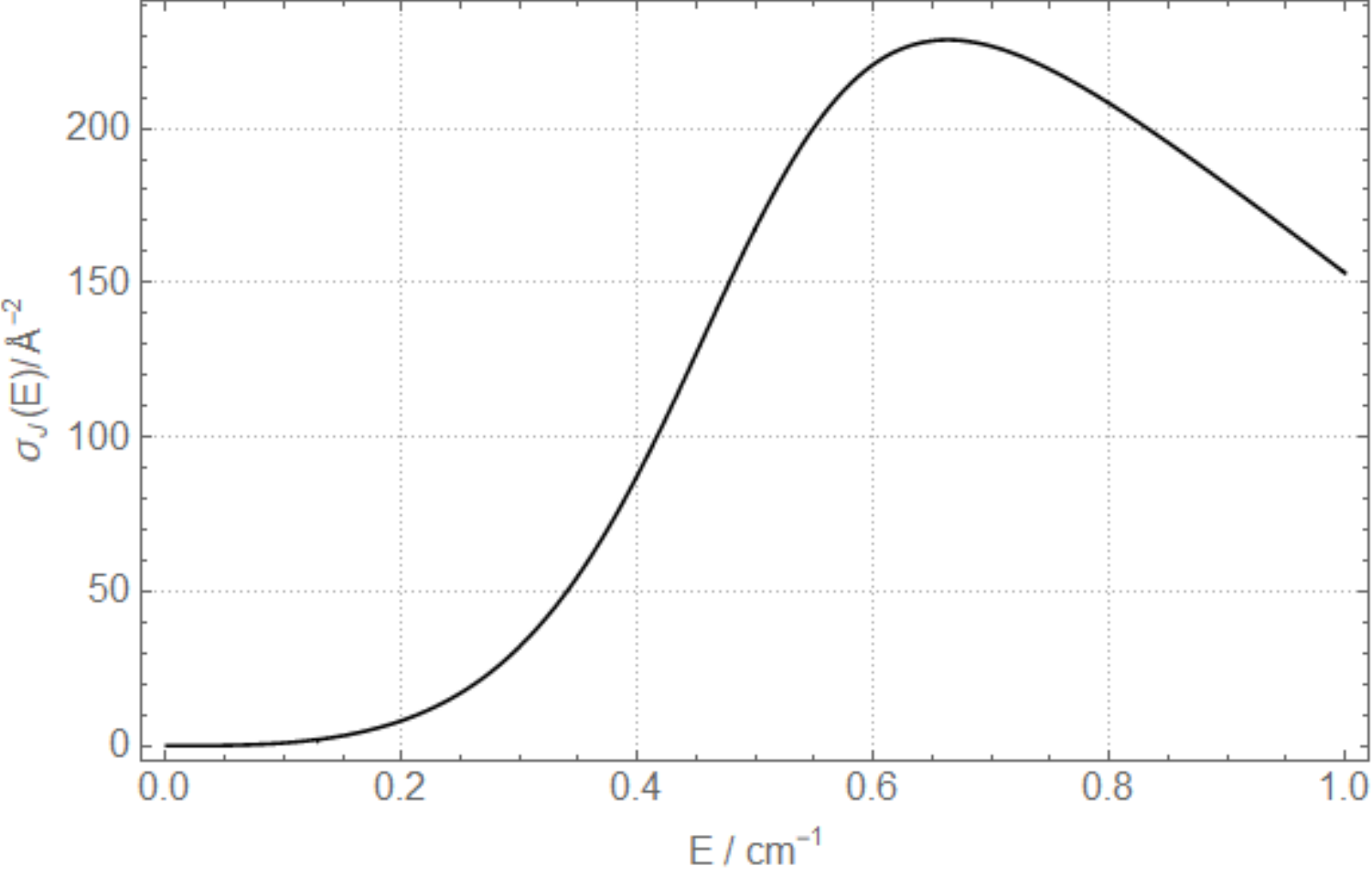}
\includegraphics[scale=0.4]{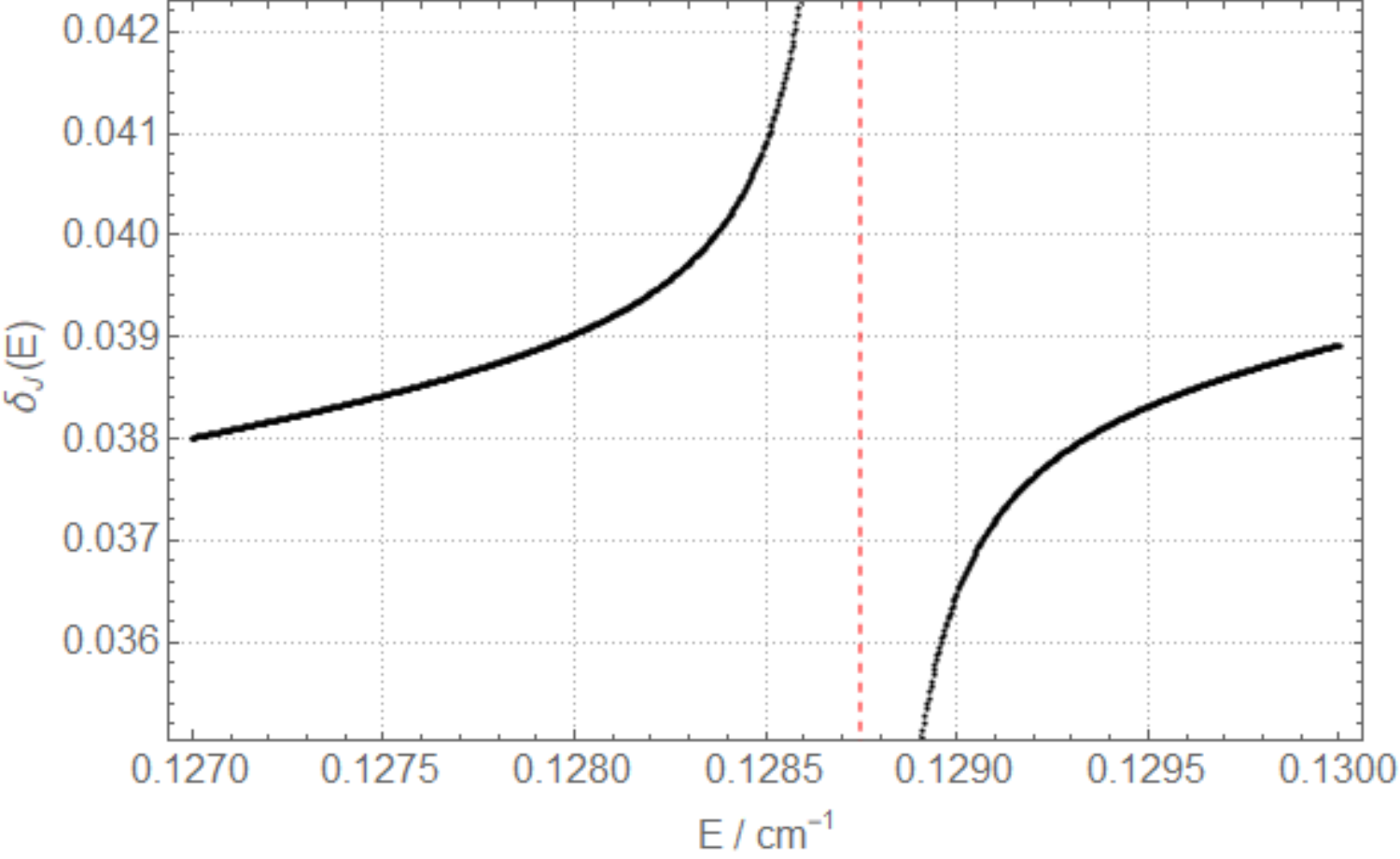}
\includegraphics[scale=0.4]{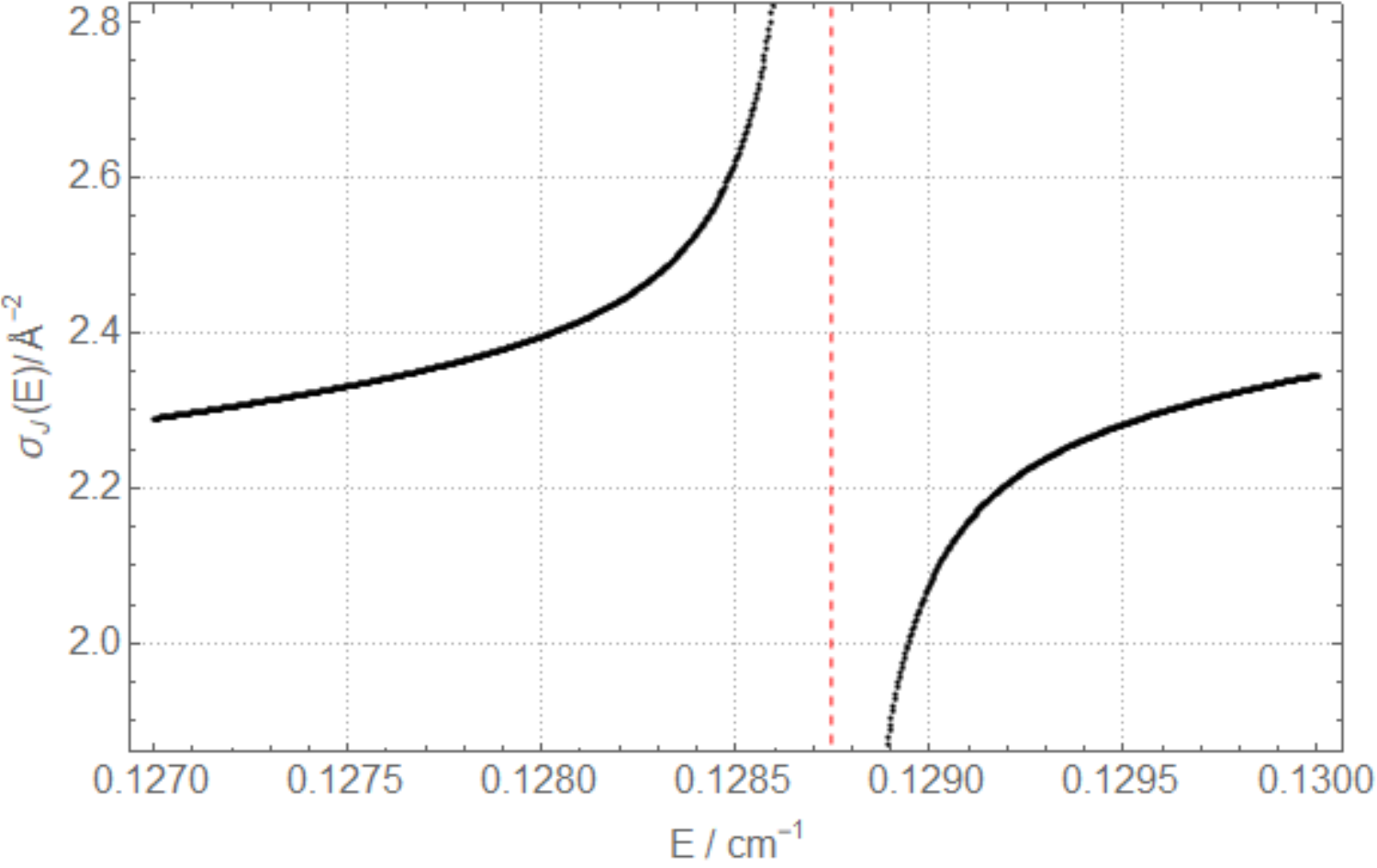}
\caption{Eigenphase (left) and cross-section (right) for the $J=9$ partial wave. 
Although the plot appears to be smooth on the scale in the top two plots, the bottom two plots are on a much narrower scale,
and show clear Fano profiles  \cite{61Fano.Rmat} associated with a resonance (position given by the dashed red line). 
Both this narrow width and its position are in agreement with the quasibound state of Myatt \etal \cite{18MyDhCh.Ar2} as described in Table \ref{tab:MyattDharmTable}.}
\label{fig:EPXSzoom}
\end{figure*}

Figure \ref{fig:EPXS} shows the eigenphase and cross-section generated
using the MD potential for $J=0$, $J=5$, and $J=10$. Figure
\ref{fig:EPXSzoom} shows the eigenphase and cross-section generated
using the MD potential for $J=9$. In all these cases, the eigenphase
and cross-section were calculated for energies between $E = 0.001$
\cm\ and $E = 1$ \cm .

The $J=0$ partial wave plots are included in Figure~\ref{fig:EPXS} to
indicate what a typical eigenphase and cross-section looks like for
this system when no resonances are present: in the $J=0$ cross-section
plot the cross-section sharply rises at low energies.

Myatt \etal \cite{18MyDhCh.Ar2} predicted (see Table
\ref{tab:MyattDharmTable}) that there should be quasibound states in the $J=5$, $J=9$, and $J=10$ partial waves.  These resonances can clearly be seen in our calculations (Figures~\ref{fig:EPXS} and
\ref{fig:EPXSzoom}) where their positions are marked with dashed lines.  These three states are the only quasibound states given by Myatt \etal
for $J\leq10$ and the only resonances detected in this work.

For the $J=5$ and $J=10$ resonances, the energy $E_{\rm{res}}$, width 
$\Gamma_{\rm{res}}$, and $A_0$ and $A_1$ parameters were fitted to the 
Breit-Wigner form of Equation~(\ref{eq:BreitWigner}), using the values quoted by 
Myatt \etal as the starting point of the fitting procedure. The very narrow $J=9$ resonance 
could not be fit in this way, and so the energy location of the width was 
determined by identifying where the eigenphase suddenly went from $\approx 
\frac{\pi}{2}$ to $\approx -\frac{\pi}{2}$ and identifying the two points either 
side of this jump;  $E_{\rm{res}}$
was taken as the mid-point between them. This energy was then inserted directly 
into the Breit-Wigner fit.

\begin{figure*}[htb!]
\centering
\includegraphics[scale=0.8]{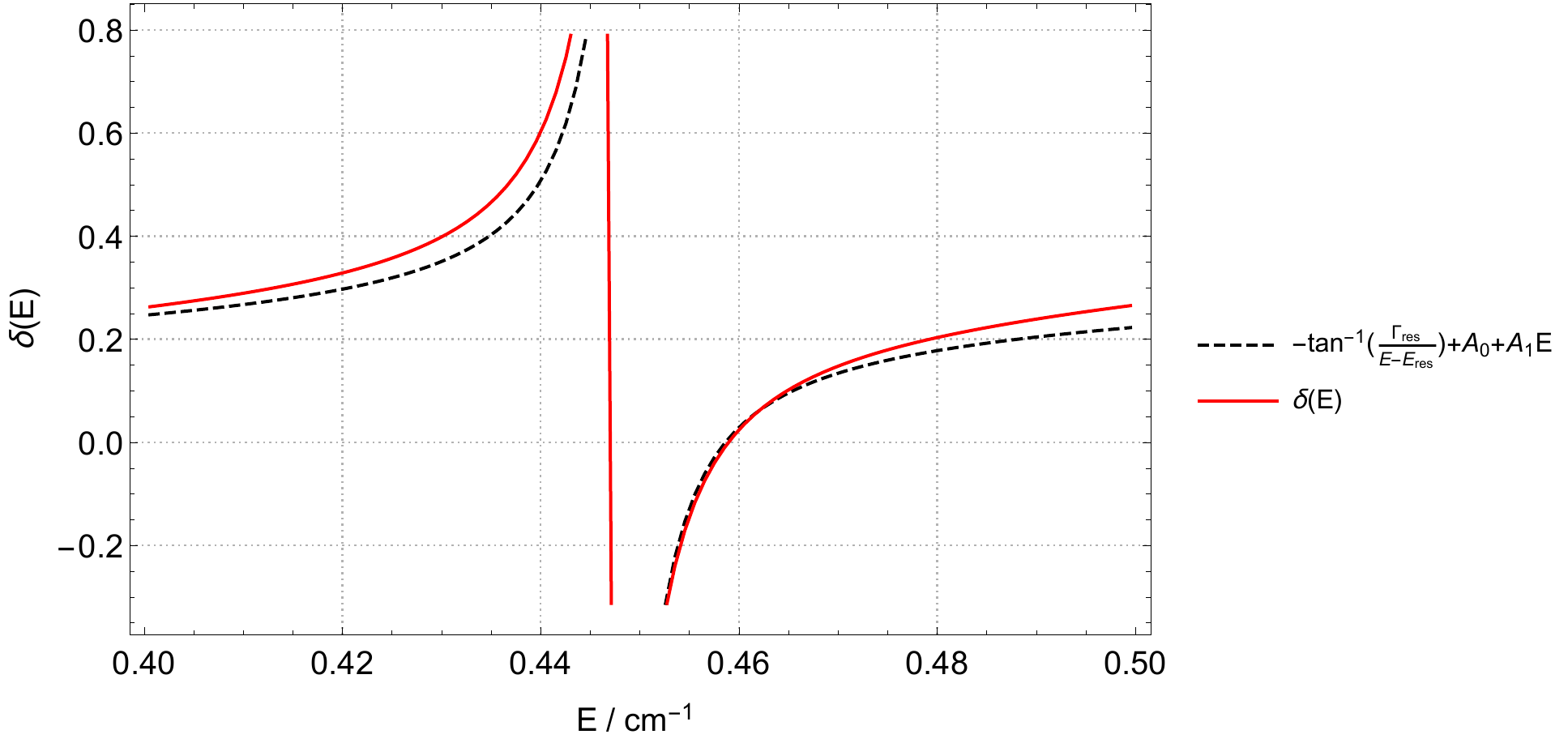}
\caption{Eigenphases in the region of the $J=10$ resonance (solid red line) with our Breit-Wigner fit (dashed black line).}
\label{fig:ResonFit}
\end{figure*}

Figure~\ref{fig:ResonFit} shows the result of this procedure for the resonance in the $J=10$ partial wave. The fitting was performed using the energy range $E = 0.4006$ \cm\ to $E = 0.499501$ \cm , using the Levenberg-Marquardt algorithm as implemented in the software Origin (OriginLab, Northhampton, MA).

Table \ref{tab:MyattDharmTable} contains the results of this fitting procedure for all three resonances studied in this work (all using the same software and algorithm with appropriate energy ranges). The narrowest resonance is for $J=9$ and there is very good agreement between our results and those quoted by Myatt \etal \cite{18MyDhCh.Ar2}. For the other two, broader resonances we find slightly different positions and widths. This is consistent with the full treatment of coupling to the continuum obtained in a scattering calculation: \textsc{LEVEL}, as used by Myatt \etal for their quasibound states, is known to be less well-adapted for characterising broader resonances \cite{16LeRoy.Rmat,jt372}. Both the resonance position and width for $J=10$ are also similar to the figures quoted by {\v{C}}{\'\i}{\v{z}}ek \etal \cite{96CiHo.Ar2}.

As Figure~\ref{fig:EPXSzoom} and Table \ref{tab:MyattDharmTable} show,
narrow resonances can be hard to detect. The only resonances detected
in this work were ones which had been previously predicted and only
needed to be corroborated. In the future, a more sophisticated resonance-detecting software such as those by Tennyson and Noble \cite{jt31} or Noble \etal \cite{93NoDoBu.Rmat}, or possibly a
procedure based on the complex analysis of the S-matrix \cite{11Burke.Rmat} such as that of {\v{C}}{\'\i}{\v{z}}ek and
Hor{\'a}{\v{c}}ek \cite{96CiHo.Rmat}, will be used to to detect
resonances which may otherwise be missed.

\begin{figure*}[htb!]
\centering
\includegraphics[scale=0.7]{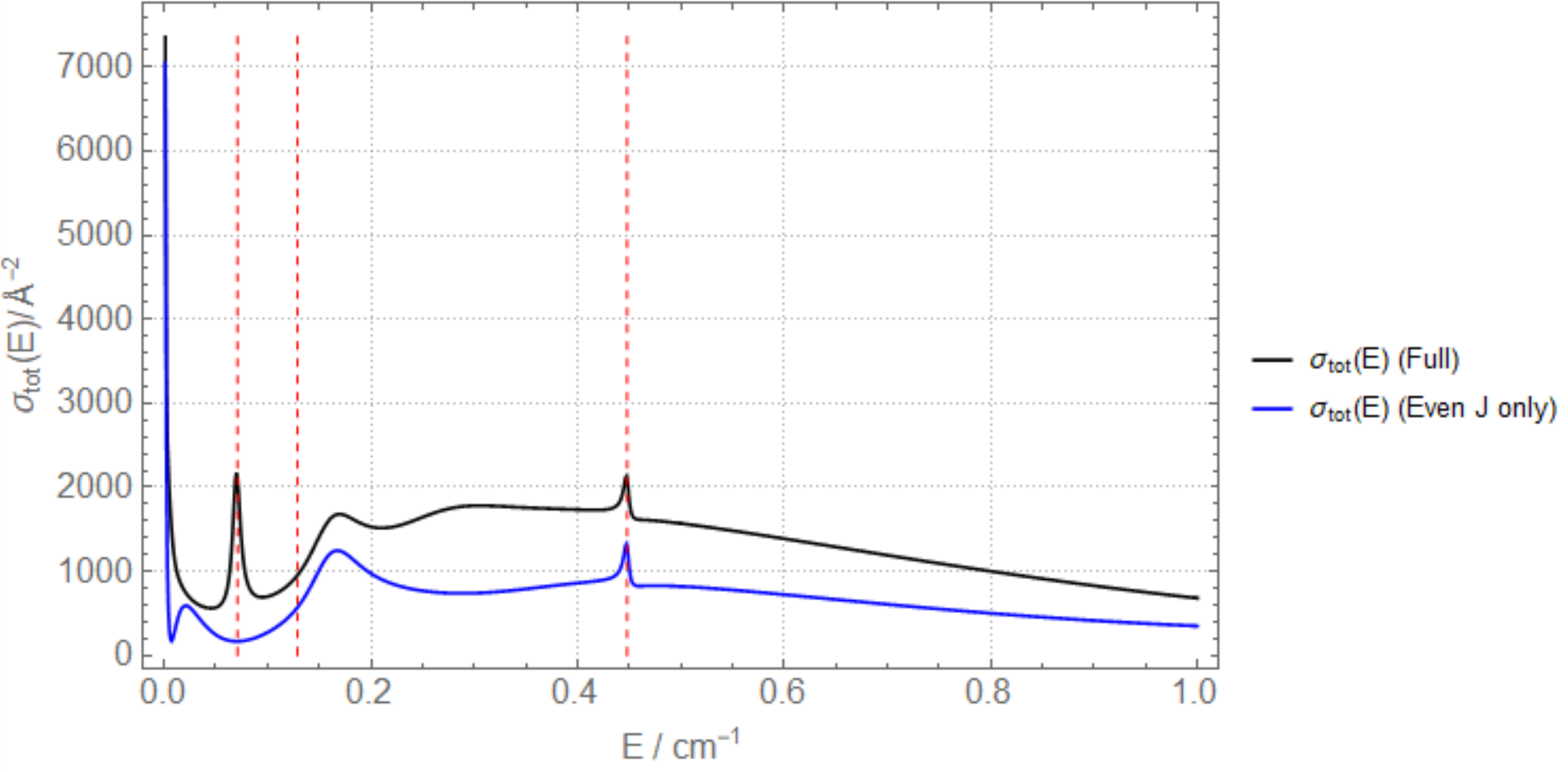}
\caption{Total cross-section when summing over the partial waves $J=0$ to 
$J=10$, using the same numerical parameters as above. The three quasibound 
states of Table \ref{tab:MyattDharmTable} are marked with dashed lines. The sum over even $J$s allows for the Pauli Principle.}
\label{fig:TotalXS}
\end{figure*}

Finally, Figure~\ref{fig:TotalXS} shows the total cross-section
generated using the RmatReact method with the MD potential. The
quasibound states predicted by Myatt \etal \cite{18MyDhCh.Ar2} are
also pictured. This figure gives a good overview of the properties of
argon-argon scattering at low energy. It is notable for having many
features. Besides the three resonances, there is also more structure
to the plot -- something that is more prevalent in heavy particle
scattering than electron-atom or electron-molecule scattering due to
the greater number of partial waves contributing to the scattering process.
Furthermore, the cross-section tends to a large value at the lowest
energies on the graph. This corroborates the feature seen in 
Figure~\ref{fig:XSlog} towards the lowest energies where the cross section 
becomes very large.

Thus far we have not considered the consequences of the Pauli principle.
$^{40}$Ar is a Boson with zero nuclear spin; as a consequence collisions with odd $J$
are forbidden. Figure~\ref{fig:TotalXS} shows the observable cross section obtained
by simply summing partial waves with even $J$. As a consequence the resonances with
$J=5$ and $J=9$ disappear and there is a pronounced Ramsauer minimum at about 0.01 \cm.

\subsection{Low-energy Scattering}
\label{LowEnergyScattering}

\begin{figure*}[htb!]
\centering
\includegraphics[scale=0.7]{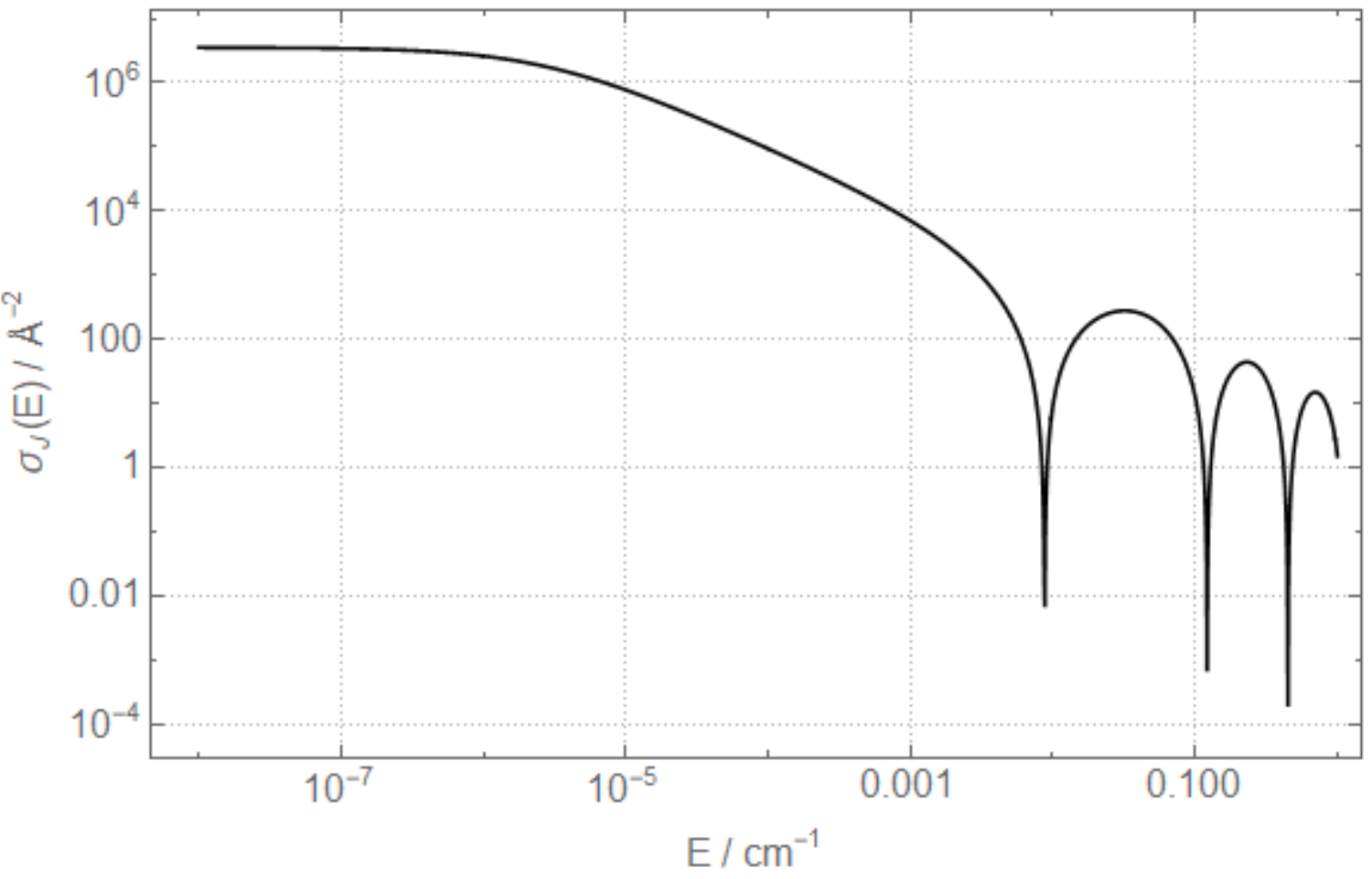}
\caption{Cross-section plot for the $J=0$ partial wave generated with the MD potential \cite{18MyDhCh.Ar2}. The plot is placed on log-log axes. At low energy the plot exhibits the signature constant scaling behaviour of low-energy scattering.}
\label{fig:XSlog}
\end{figure*}

In order to analyse low-energy scattering behaviour, the cross-section for $J=0$ was plotted for $E =10^{-8}$ \cm\ to E = 1 \cm\ on a log-log axis, see Figure~\ref{fig:XSlog}. The same numerical parameters were used as in Section \ref{Resonances}. The plot shows that the cross-section tends towards a constant at lower energies, which is predicted by Equation~(\ref{eq:lowenerg}).

\begin{figure}
\centering
\includegraphics[scale=0.75]{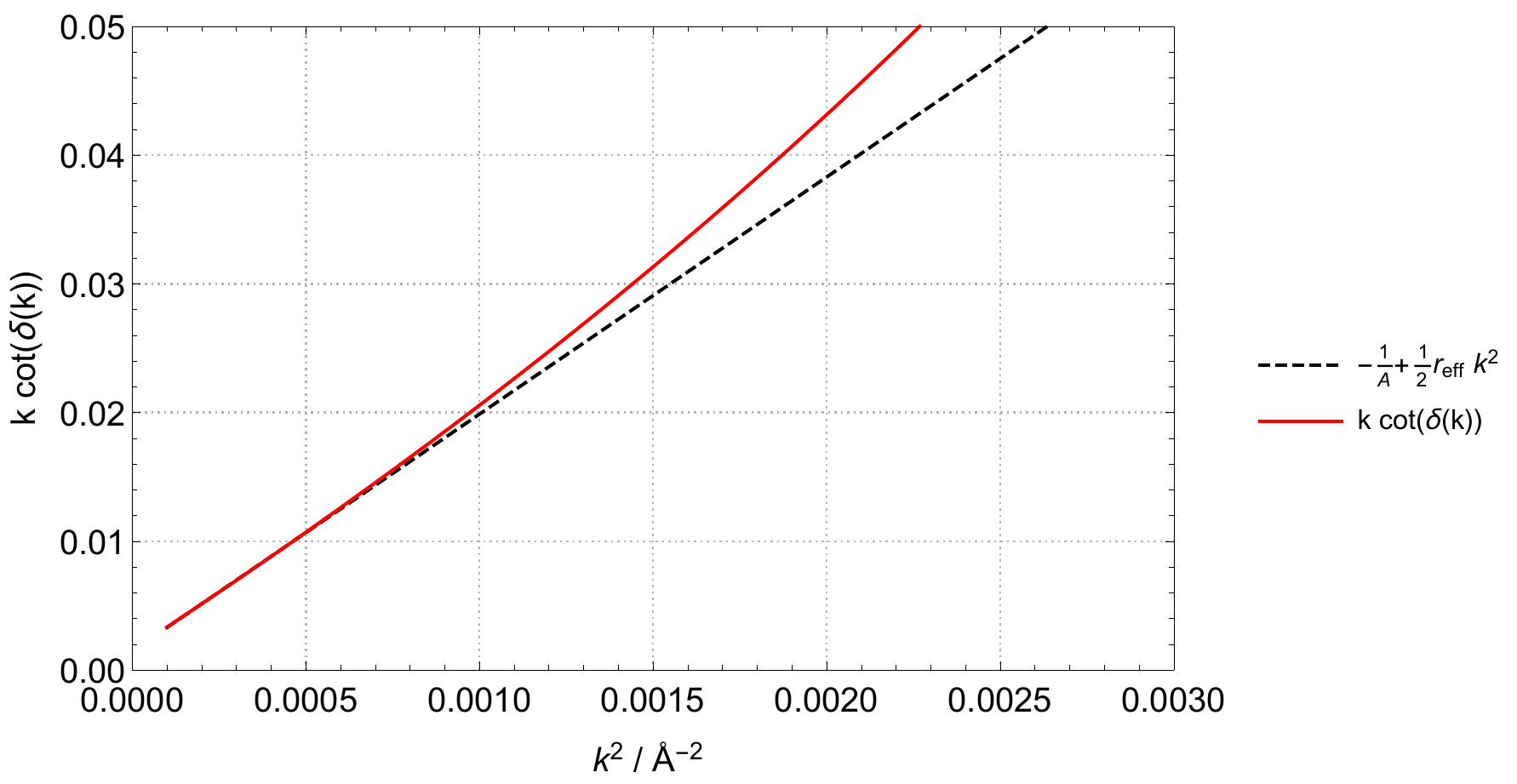}
\caption{Plot of $k \cot{\delta(k)}$ against $k^2$ for low values of $k$ using the Aziz potential \cite{93Aziz.Ar2}. }
\label{fig:BarlettaPlot}
\end{figure}

Figure~\ref{fig:BarlettaPlot} analyses the region of validity of the
low-energy linear fit of Equation~(\ref{eq:lowenerg}). It is designed to
re-create a plot shared as private communications by the authors of Ref.
\cite{jt486}.

The solid, red line of Figure~\ref{fig:BarlettaPlot} represents the eigenphase 
calculation generated by the RmatReact method, using an R-matrix inner region 
ranging from $r_{\rm{min}} = 2.5$ \AA\ to $a_0 = 82.5$ \AA , an integration over 
1600 Lobatto grid points, and an R-matrix propagation from $a_0 = 82.5$ \AA\ to 
$a_p = 165$ \AA , with
1000 propagation iterations. The dashed line represents
Equation~(\ref{eq:lowenerg}), with the parameters $A$ and \reff\ determined
by using a least-squares linear fit of the lower-energy portion of the
red line (intercept $ = 0.00146$ \AA$^{-1}$, slope $ = 18.42$ \AA ), again using 
the software Origin. As with Table \ref{tab:BarlettaTable}, this Figure is in 
agreement with results provided in private communications by
Barletta \etal \cite{jt486}, who also computed the scattering length
of the \Art\ collision based on the potential due to Aziz
\cite{93Aziz.Ar2}. It can be seen from Figure~\ref{fig:BarlettaPlot} that the plot is only linear at 
a very low energy. 

A similar low-energy fitting procedure in Origin was performed for all five PECs studied. The values of $A$ and \reff\ were calculated in this work for the four PECs where corresponding literature values could be obtained, the comparison of which can be seen in Table \ref{tab:BarlettaTable}.

The effective ranges featured in Table \ref{tab:BarlettaTable} all
appear to be in broad agreement. This is to be expected since this
quantity is not especially sensitive to fine changes to the quantity
of the potential, and is not affected significantly by the number of
bound states \cite{11Burke.Rmat}.

The values obtained for the scattering length are found to be  sensitive 
to the energy range used in the fitting procedure, and so whilst numbers are 
quoted in Table \ref{tab:BarlettaTable}, it should be noted that these numbers 
are not intended to be definitive. 
When using the energy range of Figure \ref{fig:BarlettaPlot} for the 
low-energy fit, it is possible to obtain the scattering lengths quoted in 
\cite{jt486} to within a 5\% relative difference. However, when using a much 
lower energy range for the fit of $k^2 \approx 10^{-10}$ \AA$^{-2}$ to $k^2 
\approx 10^{-8}$ \AA$^{-2}$, the scattering lengths change significantly. (The 
effective ranges also change slightly, but are still in agreement.)
The values quoted in Table \ref{tab:BarlettaTable} are the ones created using 
the lower energy range fit. As Figure \ref{fig:BarlettaPlot} shows, this lower range is 
where the expansion of Equation(\ref{eq:lowenerg}) is most appropriate. 

The features seen towards the right of Figure \ref{fig:XSlog} correspond to 
energies where the eigenphase pass through zero. On a log-log plot of the 
cross-section these crossings manifest as the dips seen in the Figure.

Although the scattering length values diverge from each other very 
significantly, the RmatReact method was able to qualitatively corroborate each 
one.
The PECs in Table \ref{tab:BarlettaTable} which have a negative scattering 
length correspond to PECs for which there are eight bound states in literature 
(see Table \ref{tab:PECs}). The only PEC considered which supports nine bound 
states, PM \cite{05PaMuFo.Ar2}, has a large, positive scattering length. 

This is in line with the observation that the scattering length is strongly affected by 
the energy of the highest bound state. If 
the scattering length is plotted as a function of $V_{\rm{min}}$, the minimum of 
the potential, then there is a pole at points where the number of bound states 
increments by one, going up to positive infinity in one direction and down to 
negative infinity in the other \cite{11Burke.Rmat}. That means that either side 
of this pole, the scattering length can be very different: any real number is a 
potentially valid scattering length.

\begin{table}[]
\caption{Scattering lengths ($A$) and effective ranges (\reff) generated using four potentials compared to previous values. For the first three potentials, \cite{93Aziz.Ar2,05PaMuFo.Ar2,03TaTo.Ar2}, the scattering lengths and effective ranges cited are from Barletta \etal \cite{jt486}. For the fourth potential, the potential and scattering length are from the same source: Myatt \etal \cite{18MyDhCh.Ar2}.}

\begin{tabular}{lcccc}
\hline
Potential                                                    & \begin{tabular}[c]{@{}c@{}}$A$/\AA\\  (literature)  \end{tabular} & \begin{tabular}[c]{@{}c@{}}$A$/\AA \\  (this work)  \end{tabular} & \begin{tabular}[c]{@{}c@{}}\reff\ /\AA \\  (literature)  \end{tabular} & \begin{tabular}[c]{@{}c@{}}\reff\ /\AA \\  (this work)  \end{tabular}  \\ \hline
Aziz \cite{93Aziz.Ar2} & -505.6                                                                      & -647.1 & 35.94                                                                      & 35.53                                                                     \\ 
PM \cite{05PaMuFo.Ar2}& 1285 & 844.0 & 33.87                                                                      & 33.53 \\ 
TT \cite{03TaTo.Ar2}    & -60.79                                                                       & -62.50 & 50.12                                                                      & 49.20                                                                      \\ 
MD \cite{18MyDhCh.Ar2} & -714                                                                         & -709.3                                                                        & --                                                                         & 35.41                                                                       \\ \hline
\end{tabular}
\label{tab:BarlettaTable}
\end{table}

It is known \cite{00FaVaSn.Ar2,05PaMuFo.Ar2} that relativistic and nonadiabatic 
effects can impact potential parameters such as the depth of the potential. The 
different PECs studied in this work all incorporate these effects to different 
degrees. Whilst this work attempts to verify the scattering observables produced 
using these potentials, no attempt is made to assess the quality of each 
potential relative to the other ones. These effects, along with the other 
sources of uncertainty related to the PECs, are by far the biggest source of 
uncertainty and error in the results, and contribute much larger error to the 
numbers quoted here than numerical errors in the algorithm itself.

No previous values are available for the scattering length and 
effective range of the PS PEC \cite{10PaSz.Ar2}; the scattering length and 
effective range were calculated, using the same lower energy range fitting 
as the results in Table \ref{tab:BarlettaTable}. 
The scattering length was found to be 1669~\AA\ to four significant figures. 
This is noteworthy because both this work and Sahraeian \etal \cite{18SaHa.Ar2} 
claim to have detected nine bound states for this system, and so the PS PEC 
continues the pattern of large, positive scattering lengths for \Art\ PECs with 
nine bound states, as seen in Table \ref{tab:BarlettaTable}. Finally, the 
effective range was found to be 33.82~\AA , in good agreement with most of the 
other effective ranges cited in the literature and this work.

\section{Conclusions and outlook}
\label{Conclusions}

In this paper the validity and accuracy of the RmatReact method for the 
single-channel, diatomic case was demonstrated by comparing results generated 
using it to other literature results. In doing so,
the accuracy of the scattering length, and the positions of the resonances 
generated by Myatt \etal \cite{18MyDhCh.Ar2} were confirmed. Most of the widths 
of the resonances generated by Myatt \etal were also confirmed.

This paper corroborated the qualitative features of the highly divergent 
scattering lengths quoted in
Barletta \etal \cite{jt486}. This has interesting
implications for the study of the low-energy behaviour of the argon-argon 
scattering interaction -- the debate over the scattering length remains 
unresolved. 
Novel experimental techniques such as those in \cite{18BeMe.Rmat} may help to 
resolve the dispute over the \Art\ scattering length and the alleged ninth bound 
state. 

Further study of the single-channel, atom-atom scattering problem is intended. A 
resonance finder will be useful for detecting any narrow resonances missed by 
other authors. The S-matrix can be used for this purpose, and also for the 
equally useful purpose of detecting weakly-bound bound states \cite{11Burke.Rmat,jt106}.

In resolving the numerical difficulties of adapting pre-existing codes to the 
`harness' of the RmatReact method, this work paves the way for the study of more 
complex interactions with the method. Other
follow-ups to this work will include a study of a multichannel collision between 
atoms, and collisions between an atom and a diatom.

Eventually the RmatReact method is intended to evolve into a method that can be 
applied to even more complex reactants and reactions, to resolve the many 
outstanding questions in the field of ultracold scattering. A formulation of the
method for treating chemical reactions in three particle systems has recently
been presented \cite{jt758}.

\section*{Acknowledgements}
This project has received funding from the European Union's Horizon 2020 research and innovation programme under the Marie Sklodowska-Curie grant agreement No 701962, and from the EPSRC under grants EP/M507970/1 and EP/R029342/1. 
We would also like to acknowledge the contributions of the late Professor Robert J. Le Roy, in this and many other works we have been involved in.
\bibliographystyle{apsrev}


\begin{thebibliography}{86}
\expandafter\ifx\csname natexlab\endcsname\relax\def\natexlab#1{#1}\fi
\expandafter\ifx\csname bibnamefont\endcsname\relax
  \def\bibnamefont#1{#1}\fi
\expandafter\ifx\csname bibfnamefont\endcsname\relax
  \def\bibfnamefont#1{#1}\fi
\expandafter\ifx\csname citenamefont\endcsname\relax
  \def\citenamefont#1{#1}\fi
\expandafter\ifx\csname url\endcsname\relax
  \def\url#1{\texttt{#1}}\fi
\expandafter\ifx\csname urlprefix\endcsname\relax\def\urlprefix{URL }\fi
\providecommand{\bibinfo}[2]{#2}
\providecommand{\eprint}[2][]{\url{#2}}

\bibitem[{\citenamefont{Stuhl et~al.}(2014)\citenamefont{Stuhl, Hummon, and
  Ye}}]{14StHuYe.Rmat}
\bibinfo{author}{\bibfnamefont{B.~K.} \bibnamefont{Stuhl}},
  \bibinfo{author}{\bibfnamefont{M.~T.} \bibnamefont{Hummon}},
  \bibnamefont{and} \bibinfo{author}{\bibfnamefont{J.}~\bibnamefont{Ye}},
  \bibinfo{journal}{Ann. Rev. Phys. Chem.} \textbf{\bibinfo{volume}{65}},
  \bibinfo{pages}{501} (\bibinfo{year}{2014}).

\bibitem[{\citenamefont{Stuhl et~al.}(2012)\citenamefont{Stuhl, Hummon, Yeo,
  Quemener, Bohn, and Ye}}]{12StHuYe.Rmat}
\bibinfo{author}{\bibfnamefont{B.~K.} \bibnamefont{Stuhl}},
  \bibinfo{author}{\bibfnamefont{M.~T.} \bibnamefont{Hummon}},
  \bibinfo{author}{\bibfnamefont{M.}~\bibnamefont{Yeo}},
  \bibinfo{author}{\bibfnamefont{G.}~\bibnamefont{Quemener}},
  \bibinfo{author}{\bibfnamefont{J.~L.} \bibnamefont{Bohn}}, \bibnamefont{and}
  \bibinfo{author}{\bibfnamefont{J.}~\bibnamefont{Ye}},
  \bibinfo{journal}{Nature} \textbf{\bibinfo{volume}{492}},
  \bibinfo{pages}{396} (\bibinfo{year}{2012}).

\bibitem[{\citenamefont{Shuman et~al.}(2010)\citenamefont{Shuman, Barry, and
  DeMille}}]{10ShBaDe.Rmat}
\bibinfo{author}{\bibfnamefont{E.~S.} \bibnamefont{Shuman}},
  \bibinfo{author}{\bibfnamefont{J.~F.} \bibnamefont{Barry}}, \bibnamefont{and}
  \bibinfo{author}{\bibfnamefont{D.}~\bibnamefont{DeMille}},
  \bibinfo{journal}{Nature} \textbf{\bibinfo{volume}{467}},
  \bibinfo{pages}{820} (\bibinfo{year}{2010}).

\bibitem[{\citenamefont{Zhelyazkova et~al.}(2014)\citenamefont{Zhelyazkova,
  Cournol, Wall, Matsushima, Hudson, Hinds, Tarbutt, and
  Sauer}}]{14ZhCoWa.Rmat}
\bibinfo{author}{\bibfnamefont{V.}~\bibnamefont{Zhelyazkova}},
  \bibinfo{author}{\bibfnamefont{A.}~\bibnamefont{Cournol}},
  \bibinfo{author}{\bibfnamefont{T.~E.} \bibnamefont{Wall}},
  \bibinfo{author}{\bibfnamefont{A.}~\bibnamefont{Matsushima}},
  \bibinfo{author}{\bibfnamefont{J.~J.} \bibnamefont{Hudson}},
  \bibinfo{author}{\bibfnamefont{E.~A.} \bibnamefont{Hinds}},
  \bibinfo{author}{\bibfnamefont{M.~R.} \bibnamefont{Tarbutt}},
  \bibnamefont{and} \bibinfo{author}{\bibfnamefont{B.~E.} \bibnamefont{Sauer}},
  \bibinfo{journal}{Phys. Rev. A} \textbf{\bibinfo{volume}{89}},
  \bibinfo{pages}{053416} (\bibinfo{year}{2014}).

\bibitem[{\citenamefont{Molony et~al.}(2014)\citenamefont{Molony, Gregory, Ji,
  Lu, K\"oppinger, Le~Sueur, Blackley, Hutson, and Cornish}}]{14MoGrJi.Rmat}
\bibinfo{author}{\bibfnamefont{P.~K.} \bibnamefont{Molony}},
  \bibinfo{author}{\bibfnamefont{P.~D.} \bibnamefont{Gregory}},
  \bibinfo{author}{\bibfnamefont{Z.}~\bibnamefont{Ji}},
  \bibinfo{author}{\bibfnamefont{B.}~\bibnamefont{Lu}},
  \bibinfo{author}{\bibfnamefont{M.~P.} \bibnamefont{K\"oppinger}},
  \bibinfo{author}{\bibfnamefont{C.~R.} \bibnamefont{Le~Sueur}},
  \bibinfo{author}{\bibfnamefont{C.~L.} \bibnamefont{Blackley}},
  \bibinfo{author}{\bibfnamefont{J.~M.} \bibnamefont{Hutson}},
  \bibnamefont{and} \bibinfo{author}{\bibfnamefont{S.~L.}
  \bibnamefont{Cornish}}, \bibinfo{journal}{Phys. Rev. Lett.}
  \textbf{\bibinfo{volume}{113}}, \bibinfo{pages}{255301}
  (\bibinfo{year}{2014}).

\bibitem[{\citenamefont{Dawid et~al.}(2018)\citenamefont{Dawid, Lewenstein, and
  Tomza}}]{18DaLeTo.Rmat}
\bibinfo{author}{\bibfnamefont{A.}~\bibnamefont{Dawid}},
  \bibinfo{author}{\bibfnamefont{M.}~\bibnamefont{Lewenstein}},
  \bibnamefont{and} \bibinfo{author}{\bibfnamefont{M.}~\bibnamefont{Tomza}},
  \bibinfo{journal}{Phys. Rev. A} \textbf{\bibinfo{volume}{97}},
  \bibinfo{pages}{063618} (\bibinfo{year}{2018}).

\bibitem[{\citenamefont{Bohn}(2001)}]{01Bohn.Rmat}
\bibinfo{author}{\bibfnamefont{J.~L.} \bibnamefont{Bohn}},
  \bibinfo{journal}{Phys. Rev. A} \textbf{\bibinfo{volume}{63}},
  \bibinfo{pages}{052714} (\bibinfo{year}{2001}).

\bibitem[{\citenamefont{K{\"o}hler et~al.}(2006)\citenamefont{K{\"o}hler,
  G{\'o}ral, and Julienne}}]{06KoGoJu.Rmat}
\bibinfo{author}{\bibfnamefont{T.}~\bibnamefont{K{\"o}hler}},
  \bibinfo{author}{\bibfnamefont{K.}~\bibnamefont{G{\'o}ral}},
  \bibnamefont{and} \bibinfo{author}{\bibfnamefont{P.~S.}
  \bibnamefont{Julienne}}, \bibinfo{journal}{Rev. Mod. Phys.}
  \textbf{\bibinfo{volume}{78}}, \bibinfo{pages}{1311} (\bibinfo{year}{2006}).

\bibitem[{\citenamefont{Heazlewood and Softley}({2015})}]{15HeSo.Rmat}
\bibinfo{author}{\bibfnamefont{B.~R.} \bibnamefont{Heazlewood}}
  \bibnamefont{and} \bibinfo{author}{\bibfnamefont{T.~P.}
  \bibnamefont{Softley}}, \bibinfo{journal}{Ann. Rev. Phys. Chem.}
  \textbf{\bibinfo{volume}{{66}}}, \bibinfo{pages}{475}
  (\bibinfo{year}{{2015}}).

\bibitem[{\citenamefont{Bell and Softley}(2009)}]{09BeSo.Rmat}
\bibinfo{author}{\bibfnamefont{M.~T.} \bibnamefont{Bell}} \bibnamefont{and}
  \bibinfo{author}{\bibfnamefont{T.~P.} \bibnamefont{Softley}},
  \bibinfo{journal}{Mol. Phys.} \textbf{\bibinfo{volume}{107}},
  \bibinfo{pages}{99} (\bibinfo{year}{2009}).

\bibitem[{\citenamefont{Bell et~al.}({2009})\citenamefont{Bell, Gingell,
  Oldham, Softley, and Willitsch}}]{09BeGiOl.Rmat}
\bibinfo{author}{\bibfnamefont{M.~T.} \bibnamefont{Bell}},
  \bibinfo{author}{\bibfnamefont{A.~D.} \bibnamefont{Gingell}},
  \bibinfo{author}{\bibfnamefont{J.~M.} \bibnamefont{Oldham}},
  \bibinfo{author}{\bibfnamefont{T.~P.} \bibnamefont{Softley}},
  \bibnamefont{and}
  \bibinfo{author}{\bibfnamefont{S.}~\bibnamefont{Willitsch}},
  \bibinfo{journal}{Faraday Disc.} \textbf{\bibinfo{volume}{{142}}},
  \bibinfo{pages}{73} (\bibinfo{year}{{2009}}).

\bibitem[{\citenamefont{Quemener and Julienne}(2012)}]{12QuJu.Rmat}
\bibinfo{author}{\bibfnamefont{G.}~\bibnamefont{Quemener}} \bibnamefont{and}
  \bibinfo{author}{\bibfnamefont{P.~S.} \bibnamefont{Julienne}},
  \bibinfo{journal}{Chem. Rev.} \textbf{\bibinfo{volume}{112}},
  \bibinfo{pages}{4949} (\bibinfo{year}{2012}).

\bibitem[{\citenamefont{Ospelkaus et~al.}(2010)\citenamefont{Ospelkaus, Ni,
  Wang, De~Miranda, Neyenhuis, Qu{\'e}m{\'e}ner, Julienne, Bohn, Jin, and
  Ye}}]{10OsNiWa.Rmat}
\bibinfo{author}{\bibfnamefont{S.}~\bibnamefont{Ospelkaus}},
  \bibinfo{author}{\bibfnamefont{K.-K.} \bibnamefont{Ni}},
  \bibinfo{author}{\bibfnamefont{D.}~\bibnamefont{Wang}},
  \bibinfo{author}{\bibfnamefont{M.~H.~G.} \bibnamefont{De~Miranda}},
  \bibinfo{author}{\bibfnamefont{B.}~\bibnamefont{Neyenhuis}},
  \bibinfo{author}{\bibfnamefont{G.}~\bibnamefont{Qu{\'e}m{\'e}ner}},
  \bibinfo{author}{\bibfnamefont{P.~S.} \bibnamefont{Julienne}},
  \bibinfo{author}{\bibfnamefont{J.~L.} \bibnamefont{Bohn}},
  \bibinfo{author}{\bibfnamefont{D.~S.} \bibnamefont{Jin}}, \bibnamefont{and}
  \bibinfo{author}{\bibfnamefont{J.}~\bibnamefont{Ye}},
  \bibinfo{journal}{Science} \textbf{\bibinfo{volume}{327}},
  \bibinfo{pages}{853} (\bibinfo{year}{2010}).

\bibitem[{\citenamefont{Chin et~al.}(2010)\citenamefont{Chin, Grimm, Julienne,
  and Tiesinga}}]{10ChGrJu.Rmat}
\bibinfo{author}{\bibfnamefont{C.}~\bibnamefont{Chin}},
  \bibinfo{author}{\bibfnamefont{R.}~\bibnamefont{Grimm}},
  \bibinfo{author}{\bibfnamefont{P.}~\bibnamefont{Julienne}}, \bibnamefont{and}
  \bibinfo{author}{\bibfnamefont{E.}~\bibnamefont{Tiesinga}},
  \bibinfo{journal}{Rev. Mod. Phys.} \textbf{\bibinfo{volume}{82}},
  \bibinfo{pages}{1225} (\bibinfo{year}{2010}).

\bibitem[{\citenamefont{Pellegrini et~al.}(2008)\citenamefont{Pellegrini,
  Gacesa, and C{\^o}t{\'e}}}]{08PeGaCo.Rmat}
\bibinfo{author}{\bibfnamefont{P.}~\bibnamefont{Pellegrini}},
  \bibinfo{author}{\bibfnamefont{M.}~\bibnamefont{Gacesa}}, \bibnamefont{and}
  \bibinfo{author}{\bibfnamefont{R.}~\bibnamefont{C{\^o}t{\'e}}},
  \bibinfo{journal}{Phys. Rev. Lett.} \textbf{\bibinfo{volume}{101}},
  \bibinfo{pages}{053201} (\bibinfo{year}{2008}).

\bibitem[{\citenamefont{Hutson et~al.}(2009)\citenamefont{Hutson, Beyene, and
  Gonz{\'a}lez-Mart{\'\i}nez}}]{09HuBeGo.Rmat}
\bibinfo{author}{\bibfnamefont{J.~M.} \bibnamefont{Hutson}},
  \bibinfo{author}{\bibfnamefont{M.}~\bibnamefont{Beyene}}, \bibnamefont{and}
  \bibinfo{author}{\bibfnamefont{M.~L.}
  \bibnamefont{Gonz{\'a}lez-Mart{\'\i}nez}}, \bibinfo{journal}{Phys. Rev.
  Lett.} \textbf{\bibinfo{volume}{103}}, \bibinfo{pages}{163201}
  (\bibinfo{year}{2009}).

\bibitem[{\citenamefont{D'Incao et~al.}(2004)\citenamefont{D'Incao, Suno, and
  Esry}}]{04DISuEs.Rmat}
\bibinfo{author}{\bibfnamefont{J.~P.} \bibnamefont{D'Incao}},
  \bibinfo{author}{\bibfnamefont{H.}~\bibnamefont{Suno}}, \bibnamefont{and}
  \bibinfo{author}{\bibfnamefont{B.~D.} \bibnamefont{Esry}},
  \bibinfo{journal}{Phys. Rev. Lett.} \textbf{\bibinfo{volume}{93}},
  \bibinfo{pages}{123201} (\bibinfo{year}{2004}).

\bibitem[{\citenamefont{Lompe et~al.}(2010)\citenamefont{Lompe, Ottenstein,
  Serwane, Wenz, Z{\"u}rn, and Jochim}}]{10LoOtSe.Rmat}
\bibinfo{author}{\bibfnamefont{T.}~\bibnamefont{Lompe}},
  \bibinfo{author}{\bibfnamefont{T.~B.} \bibnamefont{Ottenstein}},
  \bibinfo{author}{\bibfnamefont{F.}~\bibnamefont{Serwane}},
  \bibinfo{author}{\bibfnamefont{A.~N.} \bibnamefont{Wenz}},
  \bibinfo{author}{\bibfnamefont{G.}~\bibnamefont{Z{\"u}rn}}, \bibnamefont{and}
  \bibinfo{author}{\bibfnamefont{S.}~\bibnamefont{Jochim}},
  \bibinfo{journal}{Science} \textbf{\bibinfo{volume}{330}},
  \bibinfo{pages}{940} (\bibinfo{year}{2010}).

\bibitem[{\citenamefont{Ferlaino et~al.}(2011)\citenamefont{Ferlaino, Zenesini,
  Berninger, Huang, N{\"a}gerl, and Grimm}}]{11FeZeBe.Rmat}
\bibinfo{author}{\bibfnamefont{F.}~\bibnamefont{Ferlaino}},
  \bibinfo{author}{\bibfnamefont{A.}~\bibnamefont{Zenesini}},
  \bibinfo{author}{\bibfnamefont{M.}~\bibnamefont{Berninger}},
  \bibinfo{author}{\bibfnamefont{B.}~\bibnamefont{Huang}},
  \bibinfo{author}{\bibfnamefont{H.-C.} \bibnamefont{N{\"a}gerl}},
  \bibnamefont{and} \bibinfo{author}{\bibfnamefont{R.}~\bibnamefont{Grimm}},
  \bibinfo{journal}{Few-Body Systems} \textbf{\bibinfo{volume}{51}},
  \bibinfo{pages}{113} (\bibinfo{year}{2011}).

\bibitem[{\citenamefont{Wang and Esry}(2009)}]{09WaEs.Rmat}
\bibinfo{author}{\bibfnamefont{Y.}~\bibnamefont{Wang}} \bibnamefont{and}
  \bibinfo{author}{\bibfnamefont{B.~D.} \bibnamefont{Esry}},
  \bibinfo{journal}{Phys. Rev. Lett.} \textbf{\bibinfo{volume}{102}},
  \bibinfo{pages}{133201} (\bibinfo{year}{2009}).

\bibitem[{\citenamefont{D'Incao and Esry}(2006)}]{06DIEs.Rmat}
\bibinfo{author}{\bibfnamefont{J.~P.} \bibnamefont{D'Incao}} \bibnamefont{and}
  \bibinfo{author}{\bibfnamefont{B.~D.} \bibnamefont{Esry}},
  \bibinfo{journal}{Phys. Rev. A} \textbf{\bibinfo{volume}{73}},
  \bibinfo{pages}{030703} (\bibinfo{year}{2006}).

\bibitem[{\citenamefont{Carrington et~al.}(1982)\citenamefont{Carrington,
  Buttenshaw, and Kennedy}}]{82CaBuKe.Rmat}
\bibinfo{author}{\bibfnamefont{A.}~\bibnamefont{Carrington}},
  \bibinfo{author}{\bibfnamefont{J.}~\bibnamefont{Buttenshaw}},
  \bibnamefont{and} \bibinfo{author}{\bibfnamefont{R.}~\bibnamefont{Kennedy}},
  \bibinfo{journal}{Mol. Phys.} \textbf{\bibinfo{volume}{45}},
  \bibinfo{pages}{753} (\bibinfo{year}{1982}).

\bibitem[{\citenamefont{Henderson and Tennyson}(1996)}]{jt185}
\bibinfo{author}{\bibfnamefont{J.~R.} \bibnamefont{Henderson}}
  \bibnamefont{and} \bibinfo{author}{\bibfnamefont{J.}~\bibnamefont{Tennyson}},
  \bibinfo{journal}{Mol. Phys.} \textbf{\bibinfo{volume}{89}},
  \bibinfo{pages}{953} (\bibinfo{year}{1996}).

\bibitem[{\citenamefont{Kemp et~al.}(2000)\citenamefont{Kemp, Euan~Kirk, and
  McNab}}]{00KeEuMc.Rmat}
\bibinfo{author}{\bibfnamefont{F.}~\bibnamefont{Kemp}},
  \bibinfo{author}{\bibfnamefont{C.}~\bibnamefont{Euan~Kirk}},
  \bibnamefont{and} \bibinfo{author}{\bibfnamefont{I.~R.} \bibnamefont{McNab}},
  \bibinfo{journal}{Phil. Trans. A} \textbf{\bibinfo{volume}{358}},
  \bibinfo{pages}{2403} (\bibinfo{year}{2000}).

\bibitem[{\citenamefont{Tennyson et~al.}(2017)\citenamefont{Tennyson,
  Polyansky, Zobov, Alijah, and Cs\'asz\'ar}}]{jt706}
\bibinfo{author}{\bibfnamefont{J.}~\bibnamefont{Tennyson}},
  \bibinfo{author}{\bibfnamefont{O.~L.} \bibnamefont{Polyansky}},
  \bibinfo{author}{\bibfnamefont{N.~F.} \bibnamefont{Zobov}},
  \bibinfo{author}{\bibfnamefont{A.}~\bibnamefont{Alijah}}, \bibnamefont{and}
  \bibinfo{author}{\bibfnamefont{A.~G.} \bibnamefont{Cs\'asz\'ar}},
  \bibinfo{journal}{J. Phys. B: At. Mol. Opt. Phys.}
  \textbf{\bibinfo{volume}{50}}, \bibinfo{pages}{232001}
  (\bibinfo{year}{2017}).

\bibitem[{\citenamefont{Mayle et~al.}({2013})\citenamefont{Mayle, Quemener,
  Ruzic, and Bohn}}]{13MaQuGo.Rmat}
\bibinfo{author}{\bibfnamefont{M.}~\bibnamefont{Mayle}},
  \bibinfo{author}{\bibfnamefont{G.}~\bibnamefont{Quemener}},
  \bibinfo{author}{\bibfnamefont{B.~P.} \bibnamefont{Ruzic}}, \bibnamefont{and}
  \bibinfo{author}{\bibfnamefont{J.~L.} \bibnamefont{Bohn}},
  \bibinfo{journal}{Phys. Rev. A} \textbf{\bibinfo{volume}{{87}}},
  \bibinfo{pages}{012709} (\bibinfo{year}{{2013}}).

\bibitem[{\citenamefont{Tennyson}(2010)}]{jt474}
\bibinfo{author}{\bibfnamefont{J.}~\bibnamefont{Tennyson}},
  \bibinfo{journal}{Phys. Rep.} \textbf{\bibinfo{volume}{491}},
  \bibinfo{pages}{29} (\bibinfo{year}{2010}).

\bibitem[{\citenamefont{Burke}(2011)}]{11Burke.Rmat}
\bibinfo{author}{\bibfnamefont{P.~G.} \bibnamefont{Burke}},
  \emph{\bibinfo{title}{R-Matrix Theory of Atomic Collisions: Application to
  Atomic, Molecular and Optical Processes}}, vol.~\bibinfo{volume}{61}
  (\bibinfo{publisher}{Springer Science \& Business Media},
  \bibinfo{year}{2011}).

\bibitem[{\citenamefont{Bocchetta and Gerratt}(1985)}]{85BoGe.Rmat}
\bibinfo{author}{\bibfnamefont{C.~J.} \bibnamefont{Bocchetta}}
  \bibnamefont{and} \bibinfo{author}{\bibfnamefont{J.}~\bibnamefont{Gerratt}},
  \bibinfo{journal}{J. Chem. Phys.} \textbf{\bibinfo{volume}{82}},
  \bibinfo{pages}{1351} (\bibinfo{year}{1985}).

\bibitem[{\citenamefont{Light and Walker}(1976)}]{76LiWa.Rmat}
\bibinfo{author}{\bibfnamefont{J.~C.} \bibnamefont{Light}} \bibnamefont{and}
  \bibinfo{author}{\bibfnamefont{R.~B.} \bibnamefont{Walker}},
  \bibinfo{journal}{J. Chem. Phys.} \textbf{\bibinfo{volume}{65}},
  \bibinfo{pages}{4272} (\bibinfo{year}{1976}).

\bibitem[{\citenamefont{Walker and Light}(1980)}]{80WaLi.Rmat}
\bibinfo{author}{\bibfnamefont{R.~B.} \bibnamefont{Walker}} \bibnamefont{and}
  \bibinfo{author}{\bibfnamefont{J.~C.} \bibnamefont{Light}},
  \bibinfo{journal}{Ann. Rev. Phys. Chem.} \textbf{\bibinfo{volume}{31}},
  \bibinfo{pages}{401} (\bibinfo{year}{1980}).

\bibitem[{\citenamefont{Burke et~al.}(1998)\citenamefont{Burke, Greene, and
  Bohn}}]{98BuGrBo.MQDT}
\bibinfo{author}{\bibfnamefont{J.~P.} \bibnamefont{Burke}},
  \bibinfo{author}{\bibfnamefont{C.~H.} \bibnamefont{Greene}},
  \bibnamefont{and} \bibinfo{author}{\bibfnamefont{J.~L.} \bibnamefont{Bohn}},
  \bibinfo{journal}{Phys. Rev. Lett.} \textbf{\bibinfo{volume}{81}},
  \bibinfo{pages}{3355} (\bibinfo{year}{1998}).

\bibitem[{\citenamefont{Raoult and Mies}(2004)}]{04RaMi.MQDT}
\bibinfo{author}{\bibfnamefont{M.}~\bibnamefont{Raoult}} \bibnamefont{and}
  \bibinfo{author}{\bibfnamefont{F.~H.} \bibnamefont{Mies}},
  \bibinfo{journal}{Phys. Rev. A} \textbf{\bibinfo{volume}{70}},
  \bibinfo{pages}{012710} (\bibinfo{year}{2004}).

\bibitem[{\citenamefont{Gao et~al.}(2005{\natexlab{a}})\citenamefont{Gao,
  Tiesinga, Williams, and Julienne}}]{05GaTuWi.MQDT}
\bibinfo{author}{\bibfnamefont{B.}~\bibnamefont{Gao}},
  \bibinfo{author}{\bibfnamefont{E.}~\bibnamefont{Tiesinga}},
  \bibinfo{author}{\bibfnamefont{C.~J.} \bibnamefont{Williams}},
  \bibnamefont{and} \bibinfo{author}{\bibfnamefont{P.~S.}
  \bibnamefont{Julienne}}, \bibinfo{journal}{Phys. Rev. A}
  \textbf{\bibinfo{volume}{72}}, \bibinfo{pages}{042719}
  (\bibinfo{year}{2005}{\natexlab{a}}).

\bibitem[{\citenamefont{Croft et~al.}(2011)\citenamefont{Croft, Wallis, Hutson,
  and Julienne}}]{11CrWaHu.MQDT}
\bibinfo{author}{\bibfnamefont{J.~F.~E.} \bibnamefont{Croft}},
  \bibinfo{author}{\bibfnamefont{A.~O.~G.} \bibnamefont{Wallis}},
  \bibinfo{author}{\bibfnamefont{J.~M.} \bibnamefont{Hutson}},
  \bibnamefont{and} \bibinfo{author}{\bibfnamefont{P.~S.}
  \bibnamefont{Julienne}}, \bibinfo{journal}{Phys. Rev. A}
  \textbf{\bibinfo{volume}{84}}, \bibinfo{pages}{042703}
  (\bibinfo{year}{2011}).

\bibitem[{\citenamefont{Croft et~al.}(2012)\citenamefont{Croft, Hutson, and
  Julienne}}]{12CrHuJu.MQDT}
\bibinfo{author}{\bibfnamefont{J.~F.~E.} \bibnamefont{Croft}},
  \bibinfo{author}{\bibfnamefont{J.~M.} \bibnamefont{Hutson}},
  \bibnamefont{and} \bibinfo{author}{\bibfnamefont{P.~S.}
  \bibnamefont{Julienne}}, \bibinfo{journal}{Phys. Rev. A}
  \textbf{\bibinfo{volume}{86}}, \bibinfo{pages}{022711}
  (\bibinfo{year}{2012}).

\bibitem[{\citenamefont{Tennyson et~al.}(2016)\citenamefont{Tennyson,
  McKemmish, and Rivlin}}]{jt643}
\bibinfo{author}{\bibfnamefont{J.}~\bibnamefont{Tennyson}},
  \bibinfo{author}{\bibfnamefont{L.~K.} \bibnamefont{McKemmish}},
  \bibnamefont{and} \bibinfo{author}{\bibfnamefont{T.}~\bibnamefont{Rivlin}},
  \bibinfo{journal}{Faraday Discuss.} \textbf{\bibinfo{volume}{195}},
  \bibinfo{pages}{31} (\bibinfo{year}{2016}).

\bibitem[{\citenamefont{Rivlin et~al.}(2018)\citenamefont{Rivlin, McKemmish,
  and Tennyson}}]{jt727}
\bibinfo{author}{\bibfnamefont{T.}~\bibnamefont{Rivlin}},
  \bibinfo{author}{\bibfnamefont{L.~K.} \bibnamefont{McKemmish}},
  \bibnamefont{and} \bibinfo{author}{\bibfnamefont{J.}~\bibnamefont{Tennyson}},
  in \emph{\bibinfo{booktitle}{Quantum Collisions and Confinement of Atomic and
  Molecular Species, and Photons}}, edited by
  \bibinfo{editor}{\bibfnamefont{P.~C.} \bibnamefont{Deshmukh}},
  \bibinfo{editor}{\bibfnamefont{E.}~\bibnamefont{Krishnakumar}},
  \bibinfo{editor}{\bibfnamefont{S.}~\bibnamefont{Fritzsche}},
  \bibinfo{editor}{\bibfnamefont{M.}~\bibnamefont{Krishnamurthy}},
  \bibnamefont{and} \bibinfo{editor}{\bibfnamefont{S.}~\bibnamefont{Majumder}}
  (\bibinfo{publisher}{Springer}, \bibinfo{year}{2018}), Springer Conference
  Series.

\bibitem[{\citenamefont{Patkowski et~al.}({2005})\citenamefont{Patkowski,
  Murdachaew, Fou, and Szalewicz}}]{05PaMuFo.Ar2}
\bibinfo{author}{\bibfnamefont{K.}~\bibnamefont{Patkowski}},
  \bibinfo{author}{\bibfnamefont{G.}~\bibnamefont{Murdachaew}},
  \bibinfo{author}{\bibfnamefont{C.~M.} \bibnamefont{Fou}}, \bibnamefont{and}
  \bibinfo{author}{\bibfnamefont{K.}~\bibnamefont{Szalewicz}},
  \bibinfo{journal}{Mol. Phys.} \textbf{\bibinfo{volume}{{103}}},
  \bibinfo{pages}{2031} (\bibinfo{year}{{2005}}).

\bibitem[{\citenamefont{Aziz}(1993)}]{93Aziz.Ar2}
\bibinfo{author}{\bibfnamefont{R.~A.} \bibnamefont{Aziz}}, \bibinfo{journal}{J.
  Chem. Phys.} \textbf{\bibinfo{volume}{99}}, \bibinfo{pages}{4518}
  (\bibinfo{year}{1993}).

\bibitem[{\citenamefont{Aziz and Chen}(1977)}]{77AzCh.Ar2}
\bibinfo{author}{\bibfnamefont{R.~A.} \bibnamefont{Aziz}} \bibnamefont{and}
  \bibinfo{author}{\bibfnamefont{H.~H.} \bibnamefont{Chen}},
  \bibinfo{journal}{J. Chem. Phys.} \textbf{\bibinfo{volume}{67}},
  \bibinfo{pages}{5719} (\bibinfo{year}{1977}).

\bibitem[{\citenamefont{Song et~al.}(2010)\citenamefont{Song, Wang, Wu, and
  Liu}}]{10SoWaWu.Ar2}
\bibinfo{author}{\bibfnamefont{B.}~\bibnamefont{Song}},
  \bibinfo{author}{\bibfnamefont{X.}~\bibnamefont{Wang}},
  \bibinfo{author}{\bibfnamefont{J.}~\bibnamefont{Wu}}, \bibnamefont{and}
  \bibinfo{author}{\bibfnamefont{Z.}~\bibnamefont{Liu}},
  \bibinfo{journal}{Fluid Phase Equilibria} \textbf{\bibinfo{volume}{290}},
  \bibinfo{pages}{55} (\bibinfo{year}{2010}).

\bibitem[{\citenamefont{Tang and Toennies}(2003)}]{03TaTo.Ar2}
\bibinfo{author}{\bibfnamefont{K.~T.} \bibnamefont{Tang}} \bibnamefont{and}
  \bibinfo{author}{\bibfnamefont{J.~P.} \bibnamefont{Toennies}},
  \bibinfo{journal}{J. Chem. Phys.} \textbf{\bibinfo{volume}{118}},
  \bibinfo{pages}{4976} (\bibinfo{year}{2003}).

\bibitem[{\citenamefont{Patkowski and Szalewicz}(2010)}]{10PaSz.Ar2}
\bibinfo{author}{\bibfnamefont{K.}~\bibnamefont{Patkowski}} \bibnamefont{and}
  \bibinfo{author}{\bibfnamefont{K.}~\bibnamefont{Szalewicz}},
  \bibinfo{journal}{J. Chem. Phys.} \textbf{\bibinfo{volume}{133}},
  \bibinfo{pages}{094304} (\bibinfo{year}{2010}).

\bibitem[{\citenamefont{Myatt et~al.}(2018)\citenamefont{Myatt, Dham,
  Chandrasekhar, McCourt, and Le~Roy}}]{18MyDhCh.Ar2}
\bibinfo{author}{\bibfnamefont{P.~T.} \bibnamefont{Myatt}},
  \bibinfo{author}{\bibfnamefont{A.~K.} \bibnamefont{Dham}},
  \bibinfo{author}{\bibfnamefont{P.}~\bibnamefont{Chandrasekhar}},
  \bibinfo{author}{\bibfnamefont{F.~R.~W.} \bibnamefont{McCourt}},
  \bibnamefont{and} \bibinfo{author}{\bibfnamefont{R.~J.}
  \bibnamefont{Le~Roy}}, \bibinfo{journal}{Mol. Phys.}
  \textbf{\bibinfo{volume}{116}}, \bibinfo{pages}{1} (\bibinfo{year}{2018}).

\bibitem[{\citenamefont{Edmunds and Barker}(2014)}]{14EdBa.Ar2}
\bibinfo{author}{\bibfnamefont{P.~D.} \bibnamefont{Edmunds}} \bibnamefont{and}
  \bibinfo{author}{\bibfnamefont{P.~F.} \bibnamefont{Barker}},
  \bibinfo{journal}{Phys. Rev. Lett.} \textbf{\bibinfo{volume}{113}},
  \bibinfo{pages}{183001} (\bibinfo{year}{2014}).

\bibitem[{\citenamefont{Barletta et~al.}(2010)\citenamefont{Barletta, Tennyson,
  and Barker}}]{jt486}
\bibinfo{author}{\bibfnamefont{P.}~\bibnamefont{Barletta}},
  \bibinfo{author}{\bibfnamefont{J.}~\bibnamefont{Tennyson}}, \bibnamefont{and}
  \bibinfo{author}{\bibfnamefont{P.~F.} \bibnamefont{Barker}},
  \bibinfo{journal}{New J. Phys} \textbf{\bibinfo{volume}{12}},
  \bibinfo{pages}{113002} (\bibinfo{year}{2010}).

\bibitem[{\citenamefont{Barletta et~al.}(2009)\citenamefont{Barletta, Tennyson,
  and Barker}}]{jt458}
\bibinfo{author}{\bibfnamefont{P.}~\bibnamefont{Barletta}},
  \bibinfo{author}{\bibfnamefont{J.}~\bibnamefont{Tennyson}}, \bibnamefont{and}
  \bibinfo{author}{\bibfnamefont{P.~F.} \bibnamefont{Barker}},
  \bibinfo{journal}{New J. Phys} \textbf{\bibinfo{volume}{11}},
  \bibinfo{pages}{055029} (\bibinfo{year}{2009}).

\bibitem[{\citenamefont{Slav{\'{\i}́}{\v{c}}ek
  et~al.}(2003)\citenamefont{Slav{\'{\i}́}{\v{c}}ek, Kalus, Pa{\v{s}}ka,
  Odv{\'a}rkov{\'a}, Hobza, and Malijevsk{\`y}}}]{03SlKaPa.Ar2}
\bibinfo{author}{\bibfnamefont{P.}~\bibnamefont{Slav{\'{\i}́}{\v{c}}ek}},
  \bibinfo{author}{\bibfnamefont{R.}~\bibnamefont{Kalus}},
  \bibinfo{author}{\bibfnamefont{P.}~\bibnamefont{Pa{\v{s}}ka}},
  \bibinfo{author}{\bibfnamefont{I.}~\bibnamefont{Odv{\'a}rkov{\'a}}},
  \bibinfo{author}{\bibfnamefont{P.}~\bibnamefont{Hobza}}, \bibnamefont{and}
  \bibinfo{author}{\bibfnamefont{A.}~\bibnamefont{Malijevsk{\`y}}},
  \bibinfo{journal}{J. Chem. Phys.} \textbf{\bibinfo{volume}{119}},
  \bibinfo{pages}{2102} (\bibinfo{year}{2003}).

\bibitem[{\citenamefont{Meshkov et~al.}(2011)\citenamefont{Meshkov, Stolyarov,
  and Le~Roy}}]{11MeStLe.Rmat}
\bibinfo{author}{\bibfnamefont{V.~V.} \bibnamefont{Meshkov}},
  \bibinfo{author}{\bibfnamefont{A.~V.} \bibnamefont{Stolyarov}},
  \bibnamefont{and} \bibinfo{author}{\bibfnamefont{R.~J.}
  \bibnamefont{Le~Roy}}, \bibinfo{journal}{J. Chem. Phys.}
  \textbf{\bibinfo{volume}{135}}, \bibinfo{pages}{154108}
  (\bibinfo{year}{2011}).

\bibitem[{\citenamefont{Sahraeian and Hadizadeh}(2018)}]{18SaHa.Ar2}
\bibinfo{author}{\bibfnamefont{T.}~\bibnamefont{Sahraeian}} \bibnamefont{and}
  \bibinfo{author}{\bibfnamefont{M.~R.} \bibnamefont{Hadizadeh}},
  \bibinfo{journal}{Intern. J. Quantum Chem.} p. \bibinfo{pages}{e25807}
  (\bibinfo{year}{2018}).

\bibitem[{\citenamefont{Wigner}(1946)}]{46Wigner.Rmat}
\bibinfo{author}{\bibfnamefont{E.~P.} \bibnamefont{Wigner}},
  \bibinfo{journal}{Phys. Rev.} \textbf{\bibinfo{volume}{70}},
  \bibinfo{pages}{15} (\bibinfo{year}{1946}).

\bibitem[{\citenamefont{Wigner and Eisenbud}(1947)}]{47WiEi.Rmat}
\bibinfo{author}{\bibfnamefont{E.~P.} \bibnamefont{Wigner}} \bibnamefont{and}
  \bibinfo{author}{\bibfnamefont{L.}~\bibnamefont{Eisenbud}},
  \bibinfo{journal}{Phys. Rev.} \textbf{\bibinfo{volume}{72}},
  \bibinfo{pages}{29} (\bibinfo{year}{1947}).

\bibitem[{\citenamefont{Robson}(1969)}]{69Robson.Rmat}
\bibinfo{author}{\bibfnamefont{B.}~\bibnamefont{Robson}},
  \bibinfo{journal}{Nuclear Physics A} \textbf{\bibinfo{volume}{132}},
  \bibinfo{pages}{5} (\bibinfo{year}{1969}).

\bibitem[{\citenamefont{Buttle}(1967)}]{67Buttle.Rmat}
\bibinfo{author}{\bibfnamefont{P.}~\bibnamefont{Buttle}},
  \bibinfo{journal}{Phys. Rev.} \textbf{\bibinfo{volume}{160}},
  \bibinfo{pages}{719} (\bibinfo{year}{1967}).

\bibitem[{\citenamefont{Baluja et~al.}(1982)\citenamefont{Baluja, Burke, and
  Morgan}}]{82BaBuMo.Rmat}
\bibinfo{author}{\bibfnamefont{K.}~\bibnamefont{Baluja}},
  \bibinfo{author}{\bibfnamefont{P.}~\bibnamefont{Burke}}, \bibnamefont{and}
  \bibinfo{author}{\bibfnamefont{L.}~\bibnamefont{Morgan}},
  \bibinfo{journal}{Comput. Phys. Commun.} \textbf{\bibinfo{volume}{27}},
  \bibinfo{pages}{299} (\bibinfo{year}{1982}).

\bibitem[{\citenamefont{Burke and Schey}(1962)}]{62BuSc.Rmat}
\bibinfo{author}{\bibfnamefont{P.~G.} \bibnamefont{Burke}} \bibnamefont{and}
  \bibinfo{author}{\bibfnamefont{H.~M.} \bibnamefont{Schey}},
  \bibinfo{journal}{Phys. Rev.} \textbf{\bibinfo{volume}{126}},
  \bibinfo{pages}{147} (\bibinfo{year}{1962}).

\bibitem[{\citenamefont{Gailitis}(1976)}]{76Ga.Rmat}
\bibinfo{author}{\bibfnamefont{M.}~\bibnamefont{Gailitis}},
  \bibinfo{journal}{J. Phys.B: At. Mol. Phys.} \textbf{\bibinfo{volume}{9}},
  \bibinfo{pages}{843} (\bibinfo{year}{1976}).

\bibitem[{\citenamefont{Gao}(2001)}]{01Gao.Rmat}
\bibinfo{author}{\bibfnamefont{B.}~\bibnamefont{Gao}},
  \bibinfo{journal}{Physical Review A} \textbf{\bibinfo{volume}{64}},
  \bibinfo{pages}{010701} (\bibinfo{year}{2001}).

\bibitem[{\citenamefont{Gao et~al.}(2005{\natexlab{b}})\citenamefont{Gao,
  Tiesinga, Williams, and Julienne}}]{05GaTiWi.Rmat}
\bibinfo{author}{\bibfnamefont{B.}~\bibnamefont{Gao}},
  \bibinfo{author}{\bibfnamefont{E.}~\bibnamefont{Tiesinga}},
  \bibinfo{author}{\bibfnamefont{C.~J.} \bibnamefont{Williams}},
  \bibnamefont{and} \bibinfo{author}{\bibfnamefont{P.~S.}
  \bibnamefont{Julienne}}, \bibinfo{journal}{Physical Review A}
  \textbf{\bibinfo{volume}{72}}, \bibinfo{pages}{042719}
  (\bibinfo{year}{2005}{\natexlab{b}}).

\bibitem[{\citenamefont{Yurchenko et~al.}(2016)\citenamefont{Yurchenko, Lodi,
  Tennyson, and Stolyarov}}]{jt609}
\bibinfo{author}{\bibfnamefont{S.~N.} \bibnamefont{Yurchenko}},
  \bibinfo{author}{\bibfnamefont{L.}~\bibnamefont{Lodi}},
  \bibinfo{author}{\bibfnamefont{J.}~\bibnamefont{Tennyson}}, \bibnamefont{and}
  \bibinfo{author}{\bibfnamefont{A.~V.} \bibnamefont{Stolyarov}},
  \bibinfo{journal}{Comput. Phys. Commun.} \textbf{\bibinfo{volume}{202}},
  \bibinfo{pages}{262} (\bibinfo{year}{2016}).

\bibitem[{\citenamefont{Lill et~al.}(1982)\citenamefont{Lill, Parker, and
  Light}}]{82LiPaLi.Rmat}
\bibinfo{author}{\bibfnamefont{J.}~\bibnamefont{Lill}},
  \bibinfo{author}{\bibfnamefont{G.}~\bibnamefont{Parker}}, \bibnamefont{and}
  \bibinfo{author}{\bibfnamefont{J.}~\bibnamefont{Light}},
  \bibinfo{journal}{Chemical Physics Letters} \textbf{\bibinfo{volume}{89}},
  \bibinfo{pages}{483} (\bibinfo{year}{1982}).

\bibitem[{\citenamefont{Light et~al.}(1985)\citenamefont{Light, Hamilton, and
  Lill}}]{85LiHaLi.Rmat}
\bibinfo{author}{\bibfnamefont{J.}~\bibnamefont{Light}},
  \bibinfo{author}{\bibfnamefont{I.}~\bibnamefont{Hamilton}}, \bibnamefont{and}
  \bibinfo{author}{\bibfnamefont{J.}~\bibnamefont{Lill}}, \bibinfo{journal}{J.
  Chem. Phys.} \textbf{\bibinfo{volume}{82}}, \bibinfo{pages}{1400}
  (\bibinfo{year}{1985}).

\bibitem[{\citenamefont{Manolopoulos}(1993)}]{93Manolopoulos.Rmat}
\bibinfo{author}{\bibfnamefont{D.}~\bibnamefont{Manolopoulos}}, in
  \emph{\bibinfo{booktitle}{Numerical Grid Methods and Their Application to
  Schr{\"o}dinger’s Equation}} (\bibinfo{publisher}{Springer},
  \bibinfo{year}{1993}), pp. \bibinfo{pages}{57--68}.

\bibitem[{\citenamefont{Manolopoulos and Wyatt}(1988)}]{88MaWy.Rmat}
\bibinfo{author}{\bibfnamefont{D.}~\bibnamefont{Manolopoulos}}
  \bibnamefont{and} \bibinfo{author}{\bibfnamefont{R.}~\bibnamefont{Wyatt}},
  \bibinfo{journal}{Chem. Phys. Lett.} \textbf{\bibinfo{volume}{152}},
  \bibinfo{pages}{23} (\bibinfo{year}{1988}).

\bibitem[{\citenamefont{Meyer}(1994)}]{94Meyer.Rmat}
\bibinfo{author}{\bibfnamefont{H.-D.} \bibnamefont{Meyer}},
  \bibinfo{journal}{Chemical physics letters} \textbf{\bibinfo{volume}{223}},
  \bibinfo{pages}{465} (\bibinfo{year}{1994}).

\bibitem[{\citenamefont{Manolopoulos and Wyatt}(1989)}]{89MaWy.Rmat}
\bibinfo{author}{\bibfnamefont{D.}~\bibnamefont{Manolopoulos}}
  \bibnamefont{and} \bibinfo{author}{\bibfnamefont{R.}~\bibnamefont{Wyatt}},
  \bibinfo{journal}{Chemical Physics Letters} \textbf{\bibinfo{volume}{159}},
  \bibinfo{pages}{123} (\bibinfo{year}{1989}).

\bibitem[{\citenamefont{Colbert and Miller}(1992)}]{92CoMi.Rmat}
\bibinfo{author}{\bibfnamefont{D.~T.} \bibnamefont{Colbert}} \bibnamefont{and}
  \bibinfo{author}{\bibfnamefont{W.~H.} \bibnamefont{Miller}},
  \bibinfo{journal}{J. Chem. Phys.} \textbf{\bibinfo{volume}{96}},
  \bibinfo{pages}{1982} (\bibinfo{year}{1992}).

\bibitem[{\citenamefont{Zvijac and Light}(1976)}]{76ZvLi.Rmat}
\bibinfo{author}{\bibfnamefont{D.~J.} \bibnamefont{Zvijac}} \bibnamefont{and}
  \bibinfo{author}{\bibfnamefont{J.~C.} \bibnamefont{Light}},
  \bibinfo{journal}{Chem. Phys.} \textbf{\bibinfo{volume}{12}},
  \bibinfo{pages}{237} (\bibinfo{year}{1976}).

\bibitem[{\citenamefont{Schneider and Walker}(1979)}]{79ScWa.Rmat}
\bibinfo{author}{\bibfnamefont{B.~I.} \bibnamefont{Schneider}}
  \bibnamefont{and} \bibinfo{author}{\bibfnamefont{R.~B.}
  \bibnamefont{Walker}}, \bibinfo{journal}{J. Chem. Phys.}
  \textbf{\bibinfo{volume}{70}}, \bibinfo{pages}{2466} (\bibinfo{year}{1979}).

\bibitem[{\citenamefont{Sunderland et~al.}(2002)\citenamefont{Sunderland,
  Noble, Burke, and Burke}}]{02SuNoBu.Rmat}
\bibinfo{author}{\bibfnamefont{A.~G.} \bibnamefont{Sunderland}},
  \bibinfo{author}{\bibfnamefont{C.~J.} \bibnamefont{Noble}},
  \bibinfo{author}{\bibfnamefont{V.~M.} \bibnamefont{Burke}}, \bibnamefont{and}
  \bibinfo{author}{\bibfnamefont{P.~G.} \bibnamefont{Burke}},
  \bibinfo{journal}{Comput. Phys. Commun.} \textbf{\bibinfo{volume}{145}},
  \bibinfo{pages}{311} (\bibinfo{year}{2002}).

\bibitem[{\citenamefont{Gradshteyn and Ryzhik}(2014)}]{14GrRy.Rmat}
\bibinfo{author}{\bibfnamefont{I.~S.} \bibnamefont{Gradshteyn}}
  \bibnamefont{and} \bibinfo{author}{\bibfnamefont{I.~M.}
  \bibnamefont{Ryzhik}}, \emph{\bibinfo{title}{Table of integrals, series, and
  products}} (\bibinfo{publisher}{Academic press}, \bibinfo{year}{2014}).

\bibitem[{\citenamefont{Breit and Wigner}(1936)}]{36BrWi.Rmat}
\bibinfo{author}{\bibfnamefont{G.}~\bibnamefont{Breit}} \bibnamefont{and}
  \bibinfo{author}{\bibfnamefont{E.}~\bibnamefont{Wigner}},
  \bibinfo{journal}{Phys. Rev.} \textbf{\bibinfo{volume}{49}},
  \bibinfo{pages}{519} (\bibinfo{year}{1936}).

\bibitem[{\citenamefont{Tennyson and Noble}(1984)}]{jt31}
\bibinfo{author}{\bibfnamefont{J.}~\bibnamefont{Tennyson}} \bibnamefont{and}
  \bibinfo{author}{\bibfnamefont{C.~J.} \bibnamefont{Noble}},
  \bibinfo{journal}{Comput. Phys. Commun.} \textbf{\bibinfo{volume}{33}},
  \bibinfo{pages}{421} (\bibinfo{year}{1984}).

\bibitem[{\citenamefont{Owens and {\v{S}}pirko}(2018)}]{18OwSp.Rmat}
\bibinfo{author}{\bibfnamefont{A.}~\bibnamefont{Owens}} \bibnamefont{and}
  \bibinfo{author}{\bibfnamefont{V.}~\bibnamefont{{\v{S}}pirko}},
  \bibinfo{journal}{Journal of Physics B: Atomic, Molecular and Optical
  Physics} \textbf{\bibinfo{volume}{52}}, \bibinfo{pages}{025102}
  (\bibinfo{year}{2018}).

\bibitem[{\citenamefont{Le~Roy}(2016)}]{16LeRoy.Rmat}
\bibinfo{author}{\bibfnamefont{R.~J.} \bibnamefont{Le~Roy}},
  \bibinfo{journal}{J. Quant. Spectrosc. Radiat. Transf.}
  (\bibinfo{year}{2016}).

\bibitem[{\citenamefont{Sarpal et~al.}(1991)\citenamefont{Sarpal, Branchett,
  Tennyson, and Morgan}}]{jt106}
\bibinfo{author}{\bibfnamefont{B.~K.} \bibnamefont{Sarpal}},
  \bibinfo{author}{\bibfnamefont{S.~E.} \bibnamefont{Branchett}},
  \bibinfo{author}{\bibfnamefont{J.}~\bibnamefont{Tennyson}}, \bibnamefont{and}
  \bibinfo{author}{\bibfnamefont{L.~A.} \bibnamefont{Morgan}},
  \bibinfo{journal}{J. Phys. B: At. Mol. Opt. Phys.}
  \textbf{\bibinfo{volume}{24}}, \bibinfo{pages}{3685} (\bibinfo{year}{1991}).

\bibitem[{\citenamefont{Little and Tennyson}(2013)}]{jt560}
\bibinfo{author}{\bibfnamefont{D.~A.} \bibnamefont{Little}} \bibnamefont{and}
  \bibinfo{author}{\bibfnamefont{J.}~\bibnamefont{Tennyson}},
  \bibinfo{journal}{J. Phys. B: At. Mol. Opt. Phys.}
  \textbf{\bibinfo{volume}{46}}, \bibinfo{pages}{145102}
  (\bibinfo{year}{2013}).

\bibitem[{\citenamefont{Doss et~al.}(2006)\citenamefont{Doss, Tennyson, Saenz,
  and Jonsell}}]{jt372}
\bibinfo{author}{\bibfnamefont{N.}~\bibnamefont{Doss}},
  \bibinfo{author}{\bibfnamefont{J.}~\bibnamefont{Tennyson}},
  \bibinfo{author}{\bibfnamefont{A.}~\bibnamefont{Saenz}}, \bibnamefont{and}
  \bibinfo{author}{\bibfnamefont{S.}~\bibnamefont{Jonsell}},
  \bibinfo{journal}{Phys. Rev. C} \textbf{\bibinfo{volume}{73}},
  \bibinfo{pages}{025502} (\bibinfo{year}{2006}).

\bibitem[{\citenamefont{{\v{C}}{\'\i}{\v{z}}ek and
  Hor{\'a}{\v{c}}ek}(1996)}]{96CiHo.Ar2}
\bibinfo{author}{\bibfnamefont{M.}~\bibnamefont{{\v{C}}{\'\i}{\v{z}}ek}}
  \bibnamefont{and}
  \bibinfo{author}{\bibfnamefont{J.}~\bibnamefont{Hor{\'a}{\v{c}}ek}},
  \bibinfo{journal}{Czech. J.Phys.} \textbf{\bibinfo{volume}{46}},
  \bibinfo{pages}{55} (\bibinfo{year}{1996}).

\bibitem[{\citenamefont{Noble et~al.}(1993)\citenamefont{Noble, Dorr, and
  Burke}}]{93NoDoBu.Rmat}
\bibinfo{author}{\bibfnamefont{C.}~\bibnamefont{Noble}},
  \bibinfo{author}{\bibfnamefont{M.}~\bibnamefont{Dorr}}, \bibnamefont{and}
  \bibinfo{author}{\bibfnamefont{P.}~\bibnamefont{Burke}}, \bibinfo{journal}{J.
  Phys. B: At. Mol. Opt. Phys.} \textbf{\bibinfo{volume}{26}},
  \bibinfo{pages}{2983} (\bibinfo{year}{1993}).

\bibitem[{\citenamefont{C{\'\i}zek and Hor{\'a}cek}(1996)}]{96CiHo.Rmat}
\bibinfo{author}{\bibfnamefont{M.}~\bibnamefont{C{\'\i}zek}} \bibnamefont{and}
  \bibinfo{author}{\bibfnamefont{J.}~\bibnamefont{Hor{\'a}cek}},
  \bibinfo{journal}{J. Phys. A: Math. Gen.} \textbf{\bibinfo{volume}{29}},
  \bibinfo{pages}{6325} (\bibinfo{year}{1996}).

\bibitem[{\citenamefont{Fano}(1961)}]{61Fano.Rmat}
\bibinfo{author}{\bibfnamefont{U.}~\bibnamefont{Fano}},
  \bibinfo{journal}{Physical Review} \textbf{\bibinfo{volume}{124}},
  \bibinfo{pages}{1866} (\bibinfo{year}{1961}).

\bibitem[{\citenamefont{Faas et~al.}(2000)\citenamefont{Faas, Van~Lenthe, and
  Snijders}}]{00FaVaSn.Ar2}
\bibinfo{author}{\bibfnamefont{S.}~\bibnamefont{Faas}},
  \bibinfo{author}{\bibfnamefont{J.}~\bibnamefont{Van~Lenthe}},
  \bibnamefont{and} \bibinfo{author}{\bibfnamefont{J.}~\bibnamefont{Snijders}},
  \bibinfo{journal}{Molecular Physics} \textbf{\bibinfo{volume}{98}},
  \bibinfo{pages}{1467} (\bibinfo{year}{2000}).

\bibitem[{\citenamefont{Beyer and Merkt}(2018)}]{18BeMe.Rmat}
\bibinfo{author}{\bibfnamefont{M.}~\bibnamefont{Beyer}} \bibnamefont{and}
  \bibinfo{author}{\bibfnamefont{F.}~\bibnamefont{Merkt}},
  \bibinfo{journal}{Physical Review X} \textbf{\bibinfo{volume}{8}},
  \bibinfo{pages}{031085} (\bibinfo{year}{2018}).

\bibitem[{\citenamefont{McKemmish and Tennyson}(2019)}]{jt758}
\bibinfo{author}{\bibfnamefont{L.~K.} \bibnamefont{McKemmish}}
  \bibnamefont{and} \bibinfo{author}{\bibfnamefont{J.}~\bibnamefont{Tennyson}},
  \bibinfo{journal}{Phil. Trans. Royal Soc. London A}  (\bibinfo{year}{2019}).

\end{thebibliography}

\end{document}